%% file: zdraft.tex
\documentclass[zpreprint,zbstdefault]{zeus_paper}
\usepackage[english]{babel}
\usepackage{hangcaption}
\usepackage{multirow}
\input ./LaTeX/zeus/zeus_def.tex
\input ./LaTeX/user/def.tex
\input  zdraft-cit.tex
\begin{document}
\include{zdraft-tit}

\include{auth118_out}
\include{zdraft-txt}

\include{zdraft-bbl}

\include{zdraft-tab}

\include{zdraft-fig}

%
%
\end{document}

%% file: LaTeX/zeus/zeus_def.tex
\newcommand{\Zdetdesc}{%
A detailed description of the ZEUS detector can be found 
elsewhere~\cite{zeus:1993:bluebook}. A brief outline of the 
components that are most relevant for this analysis is given
below.\xspace}

\chardef\usc=95
\chardef\til=126
\catcode`\@=11 
\DeclareRobustCommand\xdotspace{\futurelet\@let@token\@xdotspace}
\def\@xdotspace{%
  \ifx\@let@token.\else
  \ifx\@let@token\bgroup.\else
  \ifx\@let@token\egroup.\else
  \ifx\@let@token\/.\else
  \ifx\@let@token\ .\else
  \ifx\@let@token~.\else
  \ifx\@let@token!.\else
  \ifx\@let@token,.\else
  \ifx\@let@token:.\else
  \ifx\@let@token;.\else
  \ifx\@let@token?.\else
  \ifx\@let@token/.\else
  \ifx\@let@token'.\else
  \ifx\@let@token).\else
  \ifx\@let@token-.\else
  \ifx\@let@token\@xobeysp.\else
  \ifx\@let@token\space.\else
  \ifx\@let@token\@sptoken.\else
   .\space
   \fi\fi\fi\fi\fi\fi\fi\fi\fi\fi\fi\fi\fi\fi\fi\fi\fi\fi}
\catcode`\@=12 

\newcommand{\stru}[2]{%
   \relax\ifmmode\hbox{\vrule height#1 depth#2 width0pt}%
   \else\vrule height#1 depth#2 width0pt\fi}

\newcommand{\Ronum}[1]{\uppercase\expandafter{\romannumeral#1}}
\newcommand{\ronum}[1]{\expandafter{\romannumeral#1}}
\DeclareRobustCommand{\LaTeXZ}{%
  \LaTeX\kern-.05em4\kern-.1em
  {\raisebox{-0.2ex}{$\scriptstyle\text{ZEUS}$}}\xspace}

\DeclareMathAlphabet{\mathbf}{OT1}{cmr}{bx}{sl}
\newcommand{\eVdist}{\kern-0.06667em}

\newcommand{\Gev}{{\text{Ge}\eVdist\text{V\/}}}

\newcommand{\gev}{{\,\text{Ge}\eVdist\text{V\/}}}

\newcommand{\nb}{\,\text{nb}}

\newcommand{\pbi}{\,\text{pb}^{-1}}

\newcommand{\mm}{\,\text{mm}}
\newcommand{\cm}{\,\text{cm}}

\newcommand{\slashfrac}[2]{%
  \raisebox{0.5ex}{\ensuremath #1}\kern-0.12em/\kern-0.08em
  \raisebox{-.8ex}{\ensuremath #2}}

\newcommand{\sqr}[3]{%
    {\vcenter{\hrule height.#3ex\hbox{\vrule width.#2ex height#1ex
     \kern#1ex\vrule width.#3ex}\hrule height.#2ex}}}

\catcode`\@=11 
\newcommand{\parenbar}{\mathpalette\p@renb@r}
\def\p@renb@r#1#2{\vbox{%
  \ifx#1\scriptscriptstyle \dimen@.7em\dimen@ii.2em\else
  \ifx#1\scriptstyle \dimen@.8em\dimen@ii.25em\else
  \dimen@1em\dimen@ii.4em\fi\fi \offinterlineskip
  \ialign{\hfill##\hfill\cr
    \vbox{\hrule width\dimen@ii}\cr
    \noalign{\vskip-.3ex}%
    \hbox to\dimen@{$\mathchar300\hfil\mathchar301$}\cr
    \noalign{\vskip-.3ex}%
    $#1#2$\cr}}}
\catcode`\@=12 

\newcommand{\IP}{{\rm I$\kern-0.01667em$P}\xspace}

\mathchardef\qsm=63
\mathchardef\pls=43
\mathchardef\mns=512
\mathchardef\plm=518
\mathchardef\eql=61
\mathchardef\smallleft=300
\mathchardef\smallright=301
\mathchardef\les=316
\mathchardef\gre=318
\mathchardef\leq=532
\mathchardef\grq=533

\catcode`\@=11 
\newcounter{pict@width}
\newcounter{pict@height}
\newlength{\pict@scale}
\newcommand{\psfigadd}[4]{%
\setcounter{pict@width}{1*\ratio{#2+\pict@scale/2}{\pict@scale}}
\setcounter{pict@height}{1*\ratio{#3+\pict@scale/2}{\pict@scale}}
\setlength{\unitlength}{\pict@scale}
\hbox to #2{\hspace{-\fill}\begin{picture}(\thepict@width,\thepict@height)
\put(0,0){\psfig{figure=#1,width=#2,height=#3,clip=}}
\SetScale{0.283466457}
\SetWidth{1.763889}
{#4}
\end{picture}}
}
\newcounter{pict@widthfst}
\newcounter{pict@widthscd}
\newcounter{pict@widthtot}
\newcommand{\psfigaddtwo}[7]{%
\setcounter{pict@widthfst}{1*\ratio{#2+\pict@scale/2}{\pict@scale}}
\setcounter{pict@widthscd}{1*\ratio{#2+#4+\pict@scale/2}{\pict@scale}}
\setcounter{pict@widthtot}{1*\ratio{#2+#4+#6+\pict@scale/2}{\pict@scale}}
\setcounter{pict@height}{1*\ratio{#3+\pict@scale/2}{\pict@scale}}
\setlength{\unitlength}{\pict@scale}
\hbox{\hspace{-\fill}\begin{picture}(\thepict@widthtot,\thepict@height)
\put(0,0){\psfig{figure=#1,width=#2,height=#3,clip=}}
\put(\thepict@widthscd,0){\psfig{figure=#5,width=#6,height=#3,clip=}}
\SetScale{0.283466457}
\SetWidth{1.763889}
{#7}
\end{picture}}
}
\newcommand{\psfigror}[4]{%
\setcounter{pict@width}{1*\ratio{#2+\pict@scale/2}{\pict@scale}}
\setcounter{pict@height}{1*\ratio{#3+\pict@scale/2}{\pict@scale}}
\setlength{\unitlength}{\pict@scale}
\hbox{\begin{picture}(\thepict@width,\thepict@height)
\put(0,\thepict@height){\psfig{figure=#1,width=#3,height=#2,clip=,angle=270}}
\SetScale{0.283466457}
\SetWidth{1.763889}
{#4}
\end{picture}}
}
\newcommand{\psfigrol}[4]{%
\setcounter{pict@width}{1*\ratio{#2+\pict@scale/2}{\pict@scale}}
\setcounter{pict@height}{1*\ratio{#3+\pict@scale/2}{\pict@scale}}
\setlength{\unitlength}{\pict@scale}
\hbox{\begin{picture}(\thepict@width,\thepict@height)
\put(0,0){\psfig{figure=#1,width=#3,height=#2,clip=,angle=90}}
\SetScale{0.283466457}
\SetWidth{1.763889}
{#4}
\end{picture}}
}
\catcode`\@=12 
\newlength\listtextwidth

\catcode`\@=11 
\newlength{\@tabfninsert}
\newlength{\@tabfnwidth}
\newcommand{\tabfootnote}[2]{%
  \setlength{\@tabfninsert}{0.8em}
  \setlength{\@tabfnwidth}{\textwidth}
  \addtolength{\@tabfnwidth}{-\@tabfninsert}
  \addtolength{\@tabfnwidth}{-0.4em}
  \noindent\makebox[\@tabfninsert][r]{\footnotesize$^{#1}$\hfil}\hfill%
  \parbox[t]{\@tabfnwidth}{\footnotesize #2\hfill}}
\catcode`\@=12 

%% file: zdraft-cit.tex
\def\citeCTD{{\cite{%
nim:a279:290,*npps:b32:181,*nim:a338:254%
}}\xspace}
\def\citeCAL{{\cite{%
nim:a309:77,*nim:a309:101,*nim:a321:356,*nim:a336:23%
}}\xspace}

%% file: zdraft-tit.tex
\title{
Exclusive electroproduction of $J/\psi$ mesons at HERA}
\author{ZEUS Collaboration}
\prepnum{DESY 04-052}
\date{March 2004}

\abstract{
The exclusive electroproduction of $J/\psi$ mesons, $ep \rightarrow
ep~J/\psi$,
has been studied with the ZEUS detector at HERA
for virtualities of the exchanged photon in the ranges
$0.15<Q^2<0.8 \gev^2$ and $2<Q^2<100\gev^2$
using integrated luminosities of $69\pbi$ and $83\pbi$,
respectively. The photon-proton centre-of-mass energy
was in the range $30 < W < 220\gev$
and the squared four-momentum transfer at the proton vertex 
$|t|<1\gev^2$.
The cross sections and decay angular distributions
are presented as functions of $Q^2$, $W$ and $t$.
The effective parameters of the Pomeron trajectory are 
in agreement with those found in $J/\psi$ photoproduction.
The spin-density matrix elements, calculated
from the decay angular distributions, are
consistent with the hypothesis of $s$-channel helicity conservation.
The ratio of the longitudinal to transverse
cross sections, $\sigma_L/\sigma_T$,
grows with $Q^2$, whilst no dependence on $W$ or $t$ is observed.
The results are in agreement with perturbative QCD calculations
and exhibit a strong sensitivity to the gluon distribution in the proton.
}

\makezeustitle

%% file: auth118_out.tex
\def\3{\ss}                                                                                        
\pagenumbering{Roman}                                                                              
                                                   %
\begin{center}                                                                                     
{                      \Large  The ZEUS Collaboration              }                               
\end{center}                                                                                       
  S.~Chekanov,                                                                                     
  M.~Derrick,                                                                                      
  D.~Krakauer,                                                                                     
  J.H.~Loizides$^{   1}$,                                                                          
  S.~Magill,                                                                                       
  S.~Miglioranzi$^{   1}$,                                                                         
  B.~Musgrave,                                                                                     
  J.~Repond,                                                                                       
  R.~Yoshida\\                                                                                     
 {\it Argonne National Laboratory, Argonne, Illinois 60439-4815}, USA~$^{n}$                       
\par \filbreak                                                                                     
  M.C.K.~Mattingly \\                                                                              
 {\it Andrews University, Berrien Springs, Michigan 49104-0380}, USA
% ...
\newpage                                                                                           
$^{\    1}$ also affiliated with University College London, London, UK \\                          
% ...
                                                           %
                                                           %
\newpage   
                                                           %
                                                           %
\begin{tabular}[h]{rp{14cm}}                                                                       
$^{a}$ &  supported by the Natural Sciences and Engineering Research                               
          Council of Canada (NSERC) \\                                                             
$^{b}$ &  supported by the German Federal Ministry for Education and                               
          Research (BMBF), under contract numbers HZ1GUA 2, HZ1GUB 0, HZ1PDA 5, HZ1VFA 5\\         
$^{c}$ &  supported by the MINERVA Gesellschaft f\"ur Forschung GmbH, the                          
          Israel Science Foundation, the U.S.-Israel Binational Science                            
          Foundation and the Benozyio Center                                                       
          for High Energy Physics\\                                                                
$^{d}$ &  supported by the German-Israeli Foundation and the Israel Science                        
          Foundation\\                                                                             
$^{e}$ &  supported by the Italian National Institute for Nuclear Physics (INFN) \\                
$^{f}$ &  supported by the Japanese Ministry of Education, Culture,                                
          Sports, Science and Technology (MEXT) and its grants for                                 
          Scientific Research\\                                                                    
$^{g}$ &  supported by the Korean Ministry of Education and Korea Science                          
          and Engineering Foundation\\                                                             
$^{h}$ &  supported by the Netherlands Foundation for Research on Matter (FOM)\\                   
$^{i}$ &  supported by the Polish State Committee for Scientific Research,                         
          grant no. 620/E-77/SPB/DESY/P-03/DZ 117/2003-2005\\                                      
$^{j}$ &  partially supported by the German Federal Ministry for Education                         
          and Research (BMBF)\\                                                                    
$^{k}$ &  supported by RF President grant N 1685.2003.2 for the leading                            
          scientific schools and by the Russian Ministry of Industry, Science                      
          and Technology through its grant for Scientific Research on High                         
          Energy Physics\\                                                                         
$^{l}$ &  supported by the Spanish Ministry of Education and Science                               
          through funds provided by CICYT\\                                                        
$^{m}$ &  supported by the Particle Physics and Astronomy Research Council, UK\\                   
$^{n}$ &  supported by the US Department of Energy\\                                               
$^{o}$ &  supported by the US National Science Foundation\\                                        
$^{p}$ &  supported by the Polish Ministry of Scientific Research and Information                  
          Technology, grant no. 112/E-356/SPUB/DESY/P-03/DZ 116/2003-2005\\                        
$^{q}$ &  supported by the Polish State Committee for Scientific Research,                         
          grant no. 115/E-343/SPUB-M/DESY/P-03/DZ 121/2001-2002, 2 P03B 07022\\                    
\end{tabular}                                                                                      
                                                           %
                                                           %

%% file: zdraft-txt.tex
\newcommand{\pom}{I\hspace{-0.2em}P}

\pagenumbering{arabic} 
\pagestyle{plain}
\section{Introduction}\label{sec-int}
The exclusive electroproduction of light ($\rho,\omega,\phi$)
and heavy ($J/\psi,\psi^{\prime},\Upsilon$) vector mesons, 
$ep\rightarrow eVp$,
has been investigated
at HERA~\cite{epj:c6:603,
epj:c13:371,
pl:b539:25,
epj:c12:393,
epj:c10:373,
pl:b483:23,
pl:b437:432,
pl:b541:251,
epj:c24:345}.
The increased precision of the recent data allows
the study of the dependence of this
process on the different scales involved: the mass squared 
of the vector meson, $M_V^2$,
the square of the centre-of-mass energy of the photon-proton system, $W^2$,
the exchanged-photon virtuality, $Q^2$, and the four-momentum transfer squared
at the proton vertex, $t$.

Exclusive electroproduction of vector mesons involving a sufficiently large 
scale is calculable perturbatively 
because of the QCD factorisation theorem~\cite{pr:d56:2982}. 
QCD-based models of this process assume that the exchanged virtual photon,
seen from the proton rest-frame, fluctuates into a quark-antiquark pair
which interacts with the proton via the exchange of two gluons in a 
colour-singlet configuration. 
After the interaction, the $q\bar{q}$ pair becomes a bound
state. The cross section is proportional to the generalised parton
distribution functions (GPDs)~\cite{fortp:78:610,*prl:78:610,*pr:d56:5524}
of the proton, which contain information on
the momentum distributions of the partons in the proton and their
correlations. 
At the leading-order approximation in $\ln(1/x)$ and vanishing $t$,
the generalised gluon distribution can be approximated 
by the usual gluon distribution.
The gluon density is probed at
$x \simeq (Q^2+M_V^2)/W^2$ and at a scale $\mu^2 \simeq Q^2+M_V^2$~\cite{barone:2002}.
The cross section is thus expected to rise steeply with $W$, a reflection
of the steep rise of the gluon density as $x$ decreases.
%
%

Data from exclusive $\rho$ 
production~\cite{epj:c6:603,epj:c13:371,pl:b539:25} 
show that the cross section $\sigma(\gamma^* p \rightarrow \rho p)$ 
rises with $W$ as $W^{\delta}$, with $\delta$ increasing with $Q^2$
from about 0.2 at $Q^2=0$ (photoproduction)
to about 0.8 at $Q^2 \approx 30 \gev^2$.
However, in the case of exclusive $J/\psi$ production
the cross section rises steeply with $W$ even
for photoproduction~\cite{epj:c24:345}.
It is therefore interesting to investigate $J/\psi$ production
at larger values of $Q^2$. 

This paper presents measurements of the exclusive electroproduction of 
$J/\psi$ mesons. 
Cross sections are given as functions of $W$, $Q^2$ and $t$. 
The $W$ dependence is also studied as a function of $t$.
The helicity structure of the $J/\psi$ has been investigated
to test  $s$-channel helicity conservation (SCHC) and 
to extract the ratio of the cross sections 
for longitudinally ($\sigma_L$) and transversely ($\sigma_T$) polarised
virtual photons, $R=\sigma_L/\sigma_T$, as a function of $W$, $Q^2$ and $t$.
The results are compared to perturbative QCD (pQCD) model calculations.

The data cover the kinematic range $30<W<220\gev$ and $|t|<1\gev^2$  
for two ranges of photon virtuality:
$0.15<Q^2<0.8\gev^2$ (low-$Q^2$ sample) and 
$2<Q^2<100\gev^2$ (high-$Q^2$ sample).
The low-$Q^2$ sample was measured in the $e^+e^-$ decay channel
and the high-$Q^2$ sample in both $e^+e^-$ and $\mu^+\mu^-$ channels.
The low-$Q^2$ range has been measured for the first time. 
The high-$Q^2$ sample represents more than an order of magnitude  
increase in statistics
compared to the previous ZEUS results~\cite{epj:c6:603},
and extends both the $W$ and $Q^2$ ranges of the measurement.

\section{Experimental set-up}
\label{sec-exp}
The data used for this measurement were taken at the HERA $ep$ collider using
the ZEUS detector in 1998-2000. 
During this period, HERA operated with a proton energy of $920\gev$ 
and an electron\footnote
{Hereafter, both $e^+$ and $e^-$ are referred to as electrons, unless explicitly 
stated otherwise.} 
energy of $27.5\gev$. 
The data correspond to integrated luminosities of 
$69\pbi$ for the low-$Q^2$ sample and $83\pbi$ for the high-$Q^2$ sample.

\Zdetdesc

Charged particles were reconstructed in the central tracking detector 
(CTD)~\citeCTD covering the polar-angle{\footnote{
The ZEUS coordinate system is a right-handed Cartesian system, with the $Z$
axis pointing in the proton beam direction, referred to as the ``forward
direction'', and the $X$ axis pointing left towards the centre of HERA.
The coordinate origin is at the nominal interaction point.\xspace}}
region $15^\circ<\theta<164^\circ$. The transverse-momentum resolution for
full-length tracks is $\sigma(p_T)/p_T=0.0058p_T\oplus0.0065\oplus0.0014/p_T$,
with $p_T$ in $\Gev$.

The high-resolution uranium-calorimeter (CAL)~\citeCAL consists of three parts:
the forward (FCAL), the barrel (BCAL) and the rear (RCAL) calorimeters. 
Each part
is subdivided transversely into towers and longitudinally into an 
electromagnetic section (EMC) and either one (RCAL) or two (FCAL and BCAL) 
hadronic sections. The CAL covers $99.7\%$ of the total solid angle.
The energy resolution obtained from test-beam measurements
was $\sigma(E)/E=0.18/\sqrt{E}$ in the electromagnetic sections and 
$\sigma(E)/E=0.35/\sqrt{E}$  in the hadronic sections, with $E$ in $\gev$.

The forward plug calorimeter (FPC)~\cite{nim:a450:235} was a lead-scintillator
sandwich calorimeter with readout via wavelength shifter fibres.
It was installed in the beamhole of the FCAL 
and extended the pseudorapidity coverage of the
forward calorimeter from  $\eta \lesssim 4$ to $\eta \lesssim 5$.

The beampipe calorimeter (BPC)~\cite{pl:b407:432} 
was a tungsten-scintillator sampling calorimeter installed 
to measure scattered electrons at small angles,
in the range $1.15^\circ<180^\circ - \theta<2.30^\circ$. 
The energy resolution, as measured under test-beam conditions, was 
$\sigma(E)/E = 0.17/\sqrt{E}$, with $E$ in $\Gev$. 
The impact position of the scattered electron was measured with an
accuracy of about 0.5~mm.

The small-angle rear tracking detector (SRTD)~\cite{nim:a401:63} consists of
two planes of scintillator strips read out via optical fibres and 
photomultiplier tubes. It is attached to the front face of the RCAL
and covers an angular range between $4^\circ$ and $18^\circ$ 
around the beampipe. 
The SRTD provides a transverse position resolution of 
0.3~cm~\cite{epj:c24:345}.

The hadron-electron separator installed in the RCAL 
(RHES)~\cite{zeus:1993:bluebook}
consists of silicon diodes placed at a longitudinal depth of three radiation
lengths.
The RHES provides 
an electron position resolution of 0.9~cm for a single hit and 0.5~cm
if the shower spans at least two adjacent pads~\cite{nim:a277:176}.

The luminosity was determined from the rate of the bremsstrahlung process
$ep\rightarrow e\gamma p$, 
where the photon was measured with a lead-scintillator
calorimeter~\cite{acpp:b32:2025} located at $Z= -107$~m in the HERA tunnel.
\section{Kinematics and cross sections}
\label{sec-kinem}
The following kinematic variables are used
to describe exclusive $J/\psi$ production,
$e(k)p(P)\rightarrow e(k^{\prime}) J/\psi(v) p(P^{\prime})$,
where $k$, $k^{\prime}$, $P$, $P^{\prime}$ and $v$ are, respectively,  
the four-momenta of the incident electron, scattered electron, 
incident proton, scattered proton and $J/\psi$:
\begin{itemize}
\item
$Q^2 = -q^2 = -(k - k^{ \prime})^2$, the negative four-momentum squared of
the virtual photon; 
\item
$W^2=(q+P)^2$, the squared invariant mass
of the photon-proton system; 
\item
$y=(P\cdot q)/(P\cdot k)$, the fraction of the 
electron
energy transferred to the proton in the proton rest frame;
\item 
$x=Q^2/(2 P\cdot q)$, the Bjorken variable;
\item
$t = (P-P^{\prime})^2$, 
the squared four-momentum transfer  at the proton vertex.
\end{itemize}

The kinematic variables were reconstructed with the ``constrained'' 
method~\cite{epj:c6:603} which uses the momentum  of the  
$J/\psi$  and the polar and azimuthal angles of the scattered electron.

The $ep$  cross section can be expressed in terms
of the transverse, $\sigma_T$,  
and longitudinal,   
$\sigma_L$, virtual photoproduction cross sections as
\begin{equation*}
\frac{d^2\sigma^{ep\rightarrow e\;J/\psi\;p}}{dydQ^2} = 
\Gamma_T(y,Q^2) 
\left (  \sigma_T + \epsilon \sigma_L \right ),
\end{equation*}
where $\Gamma_T$ is the flux of transverse virtual photons~\cite{pr:129:1834}
and $\epsilon$ is the ratio of 
longitudinal and transverse virtual photon fluxes, given by
$\epsilon = 2(1-y)/(1+(1-y)^2)$.
In the kinematic range studied here, $\epsilon$
is in the range $0.8<\epsilon<1$, with an average value of $0.99$.

The virtual photon-proton cross section, 
$\sigma^{\gamma^* p \rightarrow J/\psi p} \equiv \sigma_T + \epsilon \sigma_L$,
can be used to evaluate the total exclusive cross section, 
$\sigma_{\rm tot}^{\gamma^* p \rightarrow J/\psi p} \equiv \sigma_T + \sigma_L$,
through the relation
\begin{equation*}
\sigma_{\rm tot}^{\gamma^* p \rightarrow J/\psi p} = \frac{1+R}{1+\epsilon R}  \sigma^{\gamma^* p \rightarrow J/\psi p},
\end{equation*}
where $R=\sigma_L/ \sigma_T$
is the ratio of the cross sections for longitudinal and transverse
photons. 
The helicity structure of the $J/\psi$ production is used to determine $R$ 
as described in Section~\ref{section-R}.

\section{Reconstruction and selection of the events}
\label{sec-eventsel}
The signature of exclusive $J/\psi$ electroproduction,
$ep\rightarrow e\: J/\psi\:p$ ,
consists of the scattered electron
and two charged leptons from the $J/\psi$ decay, 
$e^+e^-$  or $\mu ^+\mu ^-$.
The scattered proton is deflected through a small angle and escapes undetected
down the beampipe.

The events were selected online by a three-level 
trigger~\cite{trigger:1992,nim:a355:278}. 
For the low-$Q^2$ sample, the trigger~\cite{epj:c24:345} for 
$J/\psi$ photoproduction events with decay to the $e^+e^-$ final state 
was used, while
for the high-$Q^2$ sample, the trigger required a 
scattered electron in the CAL with energy greater than $4\gev$.
 
The following criteria were applied offline to reconstruct and select the 
events~\cite{thesis:tandler:2003}:
\begin{itemize}
\item 
the energy and position of the scattered electron were measured in the 
BPC for the low-$Q^2$ sample
and in the CAL for the high-$Q^2$ sample.
The energy was required to satisfy $E_e>10 \gev$. 
The position measurement of the CAL was improved using 
the SRTD (88\% of the events) and the RHES (10\% of the events).
To ensure full containment of the electromagnetic shower, 
fiducial cuts were applied to the impact position of the electron on 
face of the RCAL;
\item
the $J/\psi$ mesons were reconstructed from the decay leptons.
Two tracks of opposite charge, well-reconstructed in the CTD
with $p_T>0.2\gev$, were selected (two-track events).
In the case of the electron decay channel, events were also selected
by requiring one well-reconstructed CTD track and 
one CAL energy cluster~\cite{briskin:phd:1998}
not related to the track (one-track events). 
In addition:
\begin{itemize}
\item
the two-track events were required to have
the higher-momentum track matched to a calorimeter energy cluster
for which the fraction of the energy deposited in the EMC was consistent
with that of an electron or a muon;
\item
the one-track events were accepted if, in addition to the measured CTD track
associated with a CAL cluster, 
the second cluster lay in the angular range outside the CTD acceptance
with energy between 2 and $10\gev$. 
Both clusters were required to have a fraction of energy
deposited in the EMC consistent with that of an electron.
\end{itemize}
For both types of events, one additional CTD track was allowed. 
If present, this track was required to match the scattered electron. 
Events with further tracks were rejected;
\item 
the position of the reconstructed vertex was required 
to be compatible with that of an $ep$ collision;
\item 
to remove events with large initial-state radiation 
and to reduce the background from photoproduction,
the requirement $45<\delta<65 \gev$ was imposed, 
where $\delta=\sum E_i(1-\cos\theta_i)$, $E_i$ is the 
energy of the $i^{th}$ calorimeter cell, $\theta_i$ its polar angle and the
sum runs over the cells associated to the final-state leptons;
\item 
to suppress non-exclusive events,
the energy of each CAL cluster not associated to any of the final-state leptons
was required to be less than 0.3 or $0.4 \gev$, depending on the CAL
section; these thresholds were set to be above the noise level of the CAL.
To suppress further the contamination from proton-dissociative events, 
$e p\rightarrow e J/\psi Y$, 
the energy in the FPC was required to be less than $1 \gev$ and the sum of the 
energy in the FCAL cells surrounding the beamhole to be
less than $0.5 \gev$. These cuts restrict the mass of the 
proton-dissociated system, $Y$, to $M_Y \lesssim 3.0 \gev$.
\end{itemize}

Unless otherwise stated, the results are quoted in the following 
kinematic range:
$|t|<1 \gev^2$,
$30<W<220 \gev$ for the electron channel 
and $45<W<160 \gev$ for the muon channel.
The larger $W$ range for the electron channel was achieved by the inclusion
of the one-track events. 
The $Q^2$ range was $0.15<Q^2<0.8 \gev^2$ for the low-$Q^2$ sample
and  $2<Q^2<100 \gev^2$ for the high-$Q^2$ sample.

The final high-$Q^2$ sample contains 728 events in the muon channel and
955 events in the electron channel, 275 of which are reconstructed using
only one track.
The final low-$Q^2$ sample contains 137 events in the electron channel, 
16 of which are reconstructed using only one track.
The distribution of the events in the $x$-$Q^2$ plane is shown in 
Fig.~\ref{fig-scattq2x}. 

\section{Monte Carlo simulation}
\label{sec-mc}
The acceptance and the effects of the detector response were determined
using samples of Monte Carlo (MC) events. 
All generated events were passed through the standard ZEUS detector
simulation, based on the {\sc Geant}~3.13 programme~\cite{tech:cern-dd-ee-84-1},
and the ZEUS trigger simulation package.

The exclusive process $ep\rightarrow e\: J/\psi\:p$ 
was modelled using the {\sc Zeusvm}~\cite{thesis:muchorowski:1996} 
MC generator
interfaced to 
{\sc Heracles}~4.6.1~\cite{cpc:69:155-tmp-3cfb28c9,*spi:www:heracles}
to account for first-order QED radiative effects.
The $\gamma^*\;p \rightarrow J/\psi\;p$ cross section was parametrised as 
$W^{\delta}e^{-b|t|}(M^2_{J/\psi}+Q^2)^{-n}$. 
The parameter values $n=2.5(2.3)$, $\delta=0.75(0.7)$ 
and $b=4.5(4.5)\gev^{-2}$  were used to describe 
the high-$Q^2$ (low-$Q^2$) data. 
The leptonic decay of the  $J/\psi$  was  simulated 
by the {\sc Photos} programme~\cite{cpc:79:291}
which includes final-state radiation from the decay leptons. 
This generator assumes SCHC
and that the ratio of the cross sections for longitudinal and
transverse photons is $R=0.5 \cdot (Q^2/M^2_{J/\psi})$.

Proton-dissociative events, $e p\rightarrow e J/\psi Y$, 
were modelled using the generator 
{\sc Epsoft}~\cite{thesis:adamczyk:1999,*thesis:kasprzak:1994}.
The $\gamma^*\;p \rightarrow J/\psi\;Y$ cross section was parametrised as
\begin{equation}
\frac{d^2\sigma^{\gamma^* p \rightarrow J/\psi Y}}{dt~dM^2}
\propto W^{\delta}e^{-b|t|}(M^2_{J/\psi}+Q^2)^{-n}M^{-\beta}_Y
\label{eq-epsoft}
\end{equation}
with the parameters $n=2.5$, $\delta=0.75$, $b=0.81\gev^{-2}$ and 
$\beta=2.57$ chosen as described in Section~\ref{pdiss_background}.

The QED background stemming from two-photon lepton-pair production
$\gamma^* \gamma\rightarrow l^+ l^-$,
where the virtual photon originates from the electron 
vertex and the second photon is radiated off the proton,  
was simulated using the {\sc Lpair}~\cite{proc:hera:1991:1478} generator
at low $Q^2$ and the {\sc Grape-Dilepton}~1.1~\cite{cpc:136:126} generator
at high $Q^2$.
The QED-Compton-like processes
with internal photon conversion were also generated with {\sc Grape}.

\section{Extraction of the $J/\psi$ signal}
\label{sec-mass}
Figure~\ref{fig-inv_mass} shows 
the invariant-mass distributions of the muon and electron pairs,
obtained after the selection described in Section~\ref{sec-eventsel}.
The MC distributions for exclusive $J/\psi$  production 
and the QED background are also shown.
The width of the resonance is dominated by the detector resolution, which
deteriorates at low and high values of $W$.

\subsection{Non-resonant background}
\label{sec-signal}
The non-resonant background was estimated from
the MC distributions of the QED-background processes:
two-photon lepton-pair production and Compton scattering.
For the low-$Q^2$ sample, 
the normalisation of the QED-background was estimated from
a two-parameter fit of the signal and the background MC distributions
to the invariant mass spectra of the data.
For the high-$Q^2$ sample, the normalisation was
based on the known cross sections 
and the integrated luminosity of the data. 
After subtraction of the non-resonant distributions, 
the $J/\psi$ signal was determined by counting 
the events in the mass windows $2.8<M_{\mu^+\mu^-}<3.4\gev$ 
for the muon channel and $2.6<M_{e^+e^-}<3.4\gev$ for the electron channel.
The lower limit on $M_{e^+e^-}$ was chosen to
include events with reduced invariant mass due to bremsstrahlung. 
The contribution of the non-resonant background in the signal range 
is typically $22\%$ for the electron channel 
and $14\%$ for the muon channel.

For the high-$Q^2$ sample,
additional background from pions misidentified as electrons or muons
was studied using a sample of events with two tracks, 
neither of which were identified as a muon or an electron.
The contribution was $(2.7\pm0.6)\%$  for the electron channel and 
$(0.8\pm0.3)\%$ for the muon channel and was subtracted
bin-by-bin for the $t$ and decay-angle distributions
and globally for the $W$ and $Q^2$ distributions.

Events from exclusive $\psi(2S)$ production contribute to the $J/\psi$ sample
through two different decay channels:  ($i$) 
$\psi (2S)\rightarrow J/\psi + neutrals$ 
(branching ratio (23.9$\pm$1.2)\%~\cite{pr:d66:010001}), 
where the neutrals are not detected in the CAL, 
and ($ii$) $\psi (2S)\rightarrow l^+l^-$
(branching ratios $(7.3\pm0.4)\cdot 10^{-3}$ for the electron channel and 
$(7.0\pm0.9)\cdot 10^{-3}$ for the muon decay channel~\cite{pr:d66:010001}), 
because of the limited resolution in the reconstruction of the invariant mass.
The contribution from both these processes to the $J/\psi$ sample was
determined using MC samples under the assumption that 
$\sigma(\psi(2S))/\sigma(J/\psi)=0.166\pm0.013$~\cite{pl:b541:251}.
A contribution of (1.8$\pm$0.2)\% was subtracted.

\subsection{Proton-dissociative background}
\label{pdiss_background}
The remaining  source of background  consists of 
$J/\psi$ production accompanied by proton dissociation,
$ep\rightarrow e\;J/\psi\;Y$, 
where the particles from the breakup of the proton  
are not detected.

Proton-dissociative events were studied using a sample of diffractive 
events selected as described in Section~\ref{sec-eventsel}, 
with the following exceptions:
\begin{itemize}
\item the elasticity criterion 
(last criterion in Section~\ref{sec-eventsel})
was not applied to the FPC and to a region of FCAL of 
approximately 50 cm radius around the beampipe;
\item events with decay-lepton tracks at angles smaller 
than $30^{\circ}$ with respect to the outgoing proton direction
were removed to ensure a rapidity gap between 
the $J/\psi$ and the system $Y$. 
\end{itemize}
Proton-dissociative events were selected by requiring
an energy larger than $1\gev$ in the FPC.
The sample of data tagged by the FPC contained 100 events for $|t|<3 \gev^2$
in the kinematic range $45 <W< 160\gev$ and $Q^2>2\gev^2$. 
The parameters (see Eq.(\ref{eq-epsoft}))
that best describe the $Q^2$, $W$ and $t$ dependences 
are $n=2.57\pm0.09$, $\delta=0.61\pm0.40$ and $b=0.81\pm0.25\gev^{-2}$.
The MC distribution of $M^2_Y$
was tuned to describe the FPC energy distribution, yielding 
$\beta=2.57\pm 0.67$. 
The values for $n$ and $\delta$ are in agreement with those
described in Section~\ref{sec-wdep}.
The latter are more precise and were used in {\sc Epsoft}.
The values of $b$ and $\beta$ are in agreement with those found for
proton-dissociative $J/\psi$ photoproduction~\cite{epj:c24:345}.
        
The fraction of proton-dissociative events in the elastic sample,
$f_{\rm p-diss}$, was  determined from the relation
$f_{\rm p-diss}=f^{\rm data}_{\rm FPC}(1/\epsilon^{\prime} -1),$
where $f^{\rm data}_{\rm FPC}$
denotes the fraction of the proton-dissociative sample tagged by the FPC
and  $\epsilon^{\prime}=32\%$ is the FPC tagging efficiency,
estimated using {\sc Epsoft}.
The fraction of proton-dissociative events in the final sample,
averaged over $t$ for $|t|<1\gev^2$,
was $f_{\rm p-diss}=(14.2\pm 2.0({\rm stat.})^{+6.8}_{-3.6}({\rm syst.}))\%$,
independent of $W$ and $Q^2$.
The systematic uncertainty was dominated by the uncertainty on $\beta$.
The fraction increases from 4\% for $0<|t|<0.1\gev^2$ 
to 20\% for $0.2<|t|<1.0\gev^2$.
The cross sections presented in the next sessions 
were corrected for this background 
in bins of $t$, and globally in $W$ and $Q^2$.

\section{Results}
\subsection{Measurement of cross sections}
\label{sec-xsec}
In each bin of a kinematic variable,
the $ep$ cross section was extracted for each decay channel
using the formula
\begin{equation*}
\sigma^{ep\rightarrow e\;J/\psi\;p} = 
\frac{(N_{\rm data} -N_{\rm bgd})  (1 - f_{\rm p-diss} ) } {\cal A B  L},
\end{equation*}
where $N_{\rm data}$ is the number of events in the data
and $N_{\rm bgd}$ is the number 
of events from the non-resonant background (QED processes and 
pionic background) and $\psi (2S)$ production.
The overall acceptance is denoted as ${\cal A}$,
${\cal B}$ accounts for the 
$J/\psi$ decay branching ratios~\cite{pr:d66:010001}, 
$(5.93\pm0.10)\%$ and $(5.88\pm0.10)\%$ for the
electron and muon channels, respectively, 
and ${\cal L}$  is the integrated luminosity.
 
The total exclusive photon-proton cross section was calculated as
\begin{equation*}
\sigma_{\rm tot}^{\gamma^* p \rightarrow J/\psi p} 
= (1/\Phi(Q^2,W))  d^2\sigma^{ep\rightarrow e\;J/\psi\;p}/dQ^2dW,
\end{equation*}
where the effective photon flux $\Phi$~\cite{prep:15:181}
contains the corrections for bin-centring and $R$, both 
estimated from the MC simulation.
The final cross section
was the error-weighted average of the cross sections for each decay channel.

The cross sections are quoted at the QED-Born level.
The radiative corrections range from  1\% to 10\% 
(on average 5\%), depending on the kinematic region.
  
The cross sections were measured for $|t|<1\gev^2$. 
Assuming $d\sigma/dt \propto e^{-b|t|}$, with $b=4.5\gev^{-2}$, 
the correction factor needed to extrapolate to the cross section
integrated over the full $t$ range is 1.012.   
In addition, for $x>0.01$ both the acceptance and
the expected cross section are small, and the
measurement in this region therefore involves an extrapolation, made in
order to quote the measurement in bins of $W$ and $Q^2$. The
uncertainty introduced by this extrapolation, as evaluated from the
MC simulation, is negligible.
\subsection{Systematic uncertainties}
\label{sec-syst}
The systematic uncertainties on the measured cross sections were determined by
varying the selection cuts and by modifying the analysis procedure.

For the low-$Q^2$ sample, the main contribution arises from the
uncertainty of $\pm 1$ mm in the position of the BPC, leading to a
$\pm 10\%$ uncertainty in the cross section.

For the high-$Q^2$ sample, the dominant sources of uncertainty 
are as follows. The corresponding average uncertainties are given 
in parentheses:
\begin{itemize}
\item the trigger efficiency ($^{+2.8}_{-1.2}\%$);
\item the fiducial volume cut on the electron position was changed by 
$\pm 1 \cm$; the SRTD alignment was changed by $\pm 2 \mm$
along the $Y$ axis ($^{+5.5}_{-3.5}\%$). 
A maximum change of $-11\%$ was observed in the lowest $Q^2$ bin;
\item the mass window used for signal extraction was extended by $0.1\gev$ 
($\pm 1.7\%$);
\item the normalisation of the QED background was changed by 
$\pm 10\%$ ($\pm 2.4\%$); 
the maximum effect of $\pm 5\%$ was found for the lowest $t$ bin. 
\end{itemize}
The uncertainty due to the subtraction of proton-dissociative background 
has been discussed in Section~\ref{pdiss_background}.
Additional contributions come from the uncertainties 
on the integrated luminosity, $\pm 2.25\%$, 
and on the  branching ratios, $\pm 1.7\%$.
Uncertainties from 
the minimum energy requirement of the scattered electron ($\pm 0.7\%$),
the elasticity cut ($^{+0.2}_{-1.9}\%$),
the selection of the electron and muon samples ($\pm 1.2\%$) and
the dependence on the MC parametrisations ($\pm 0.7\%$) were also estimated.
The total systematic uncertainty  was determined by 
adding the individual contributions in quadrature.
The correlated and uncorrelated systematic uncertainties 
were evaluated separately and were $^{+5}_{-8}\%$ and $^{+7.4}_{-6.4}\%$, 
respectively.

\subsection{Dependence on $W$ and $Q^2$}
\label{sec-wdep}
The  cross section $\sigma_{\rm tot}^{\gamma^* p \rightarrow J/\psi p}$,
measured as a function of $W$ and $Q^2$ for $|t|<1\gev^2$,
is given in Tables~\ref{tab-Wdep} and \ref{tab-Q2dep}.
The same cross section, extrapolated to the full $t$ range, 
is shown in Fig.~\ref{fig-xsec} together with 
the H1~\cite{epj:c10:373} measurements\footnote{
In Fig.~\ref{fig-xsec}a, the H1 cross sections, 
measured at $Q^2$ values of 3.5, 10.1 and 33.6$\gev^2$, 
have been rescaled to the $Q^2$ values of 3.1, 6.8 and $16\gev^2$ 
using the $Q^2$ dependence of the data measured by H1.
The systematic uncertainties due to this extrapolation were negligible.
}
as well as the ZEUS measurement of 
exclusive $J/\psi$ photoproduction~\cite{epj:c24:345}.
The H1 measurements are systematically lower than the ZEUS data.

The functional form $\sigma\propto W^{\delta}$ was fitted to the
ZEUS data; the results of the fit  are shown 
in Fig.~\ref{fig-xsec}a and in Table~\ref{tab-slopes}.
No significant variation of $\delta$ with $Q^2$ is seen.
The mean value of $\delta$ is 
$0.73\pm 0.11({\rm stat.})^{+0.04}_{-0.08}({\rm syst.})$.
It is consistent with the values found for $J/\psi$ 
photoproduction~\cite{epj:c6:603}
and for $\rho$ electroproduction at high $Q^2$~\cite{epj:c13:371}.

The function $\sigma=\sigma_0\cdot(M^2_{J/\psi}/ (Q^2+M^2_{J/\psi}))^{n}$,
fitted to the ZEUS data including the photoproduction point, 
is shown in Fig.~\ref{fig-xsec}b.
The resulting parameters are $\sigma_0=77\pm3\nb$ and $n=2.44\pm 0.08$, 
with  $\chi^2/\rm{ndf} = 4.1/7$.
The fit, which takes both the statistical and uncorrelated systematic 
uncertainties into account, describes the data well over the full $Q^2$ range.

\subsubsection{Comparison to model predictions}
\label{sec-models}
%
Models based on QCD are able to describe exclusive vector meson production at
HERA.
In such models, in the frame where the proton is at rest,
the photon emitted from the electron fluctuates into a
$q\bar{q}$ state, this
$q\bar{q}$ pair subsequently interacts with the proton through the 
exchange of gluons in a colour-singlet configuration
and eventually forms a bound meson state. 
The transverse size of the $q\bar{q}$ pair depends
on $Q^2$ and on the quark mass; 
for $Q^2 > {\cal O}(10)\gev^2$ or for heavy quarks, it is assumed to be
considerably smaller than the size of the proton. 
At such distances, the QCD coupling is small 
and perturbation theory can be applied.
The QCD factorisation theorem for hard exclusive electroproduction 
of mesons~\cite{pr:d56:2982} predicts that,
in the limit of large $Q^2$ and fixed $x$, 
the cross section can be estimated 
from a hard interaction part calculable in pQCD,
the $q\bar{q}$ wave function of the meson 
and the generalised parton distributions 
(GPDs)~\cite{fortp:78:610,*prl:78:610,*pr:d56:5524}
which contain information about the correlations of the partons 
inside the proton and their momentum distribution.
A rapid rise in the cross section with $W$ is predicted which is
related to the fast increase of the gluon density inside the proton
at small values of $x$.
A selection of the available models is compared to the data
and discussed below.
A more complete discussion on the avalaible models
is given elsewhere~\cite{barone:2002}.

Frankfurt, Koepf and Strikman (FKS)~\cite{pr:d54:3194,*pr:d57:512} have
proposed a model based on the leading-order approximation 
$\alpha_s \ln({Q^2})$.
The usual parton distribution functions (PDFs) are used. 
The $J/\psi$ wave function is estimated in the non-relativistic approximation.

In the model of Martin, Ryskin and Teubner (MRT)~\cite{pr:d62:14022},
the calculations are also performed at the leading order, 
$\alpha_s \ln({Q^2})$.
Assuming parton-hadron duality, 
the component of the $c\bar{c}$ pair which has the correct spin-parity
for the $J/\psi$ is used instead of the $J/\psi$ wave function. 
The cross section is integrated over the $J/\psi$ mass range.
The GPDs are estimated using the conventional 
next-to-leading (NLO) gluon distributions.

Gotsman et al. (GLLMN)~\cite{acpp:b34:3255} have
presented a dipole model where
the cross section is expressed as the convolution 
of the wave function of the virtual photon, 
the dipole scattering amplitude
and the $J/\psi$ wave function.
The dipole scattering amplitude 
is estimated at leading order, $\alpha_s \ln(1/x)$,
as the solution of the Balitsky-Kovchegov~\cite{np:d60:99,pr:d60:034008}
evolution equation, 
including both the linear BFKL terms due to parton splitting 
and nonlinear terms due to recombination of partons in the 
high-density region at low $x$.
The $J/\psi$ wave function is estimated in the non-relativistic approximation.

The $W$ and $Q^2$ dependence of the cross sections 
measured by ZEUS are compared to the QCD predictions
in Fig.~\ref{fig-xsec-models}.
As the full NLO corrections have not yet been estimated, 
all the models have significant normalisation uncertainties.
Therefore the normalisation was fixed 
using the ZEUS photoproduction data at $W=90\gev$;
the different normalisation factors are indicated in the figure.  
The gluon PDFs ZEUS-S~\cite{pr:d67:12007} for MRT 
and CTEQ4L~\cite{pr:d55:1280} for FKS were used. 
The $Q^2$ dependence of $\delta$ is compared
in the insert in Fig.~\ref{fig-xsec-models}a.
All models predict a rise of the cross section with increasing $W$
and have a $Q^2$ dependence similar to that of the data.

\subsubsection{Comparison to model predictions 
for different gluon parametrisations}
The MRT model was used to test three different gluon distributions:
MRST02~\cite{epj:c23:73}, CTEQ6M~\cite{JHEP:0207:012} and 
ZEUS-S~\cite{pr:d67:12007},
obtained from NLO DGLAP analyses of structure function data.
In deriving the GPDs from the PDFs, sensitivity to the gluon distribution
at very low $x$ is introduced.
Again, the predictions were normalised to 
the ZEUS photoproduction measurement at $W=90\gev$.

Figure~\ref{fig-xsec_gluons} compares the data with the predictions.
While CTEQ6M describes the $W$ and $Q^2$ dependence of the data,
MRST02 has the wrong shape in $W$, particularly at low $Q^2$.  
ZEUS-S describes the $W$ dependence but falls too quickly 
with increasing $Q^2$.

The data exhibit a strong sensitivity to the gluon distribution in the proton.
However, full NLO calculations are needed
in order to use these data in global fits to constrain the gluon density.

\subsection{Dependence on $t$}
\label{sec-tdep}
The  differential cross section, $d\sigma^{\gamma^* p \rightarrow J/\psi p} /dt$, 
measured as a function of $t$ in the range $|t|< 1\gev^2$,
is shown in Table~\ref{tab-tdep} and Fig.~\ref{fig-dsigmadt}a-d
for the high $Q^2$ sample as well as for three $Q^2$ intervals.  
A function of the form $d\sigma /dt = d\sigma /dt|_{t=0} \cdot e^{-b|t|}$  
was fitted to the data and
the results of the fit are given in Table~\ref{tab-slopes}. 
The slope parameter $b$ is shown in Fig.~\ref{fig-dsigmadt}e
as a function of $Q^2$ and is
compared to the ZEUS photoproduction and H1 electroproduction values.
No significant $Q^2$ dependence in $b$ is seen over the measured range of $Q^2$.
This behaviour is different from that of
exclusive $\rho$ electroproduction,
where the $b$ slope strongly decreases with increasing $Q^2$,
reaching the value of that of the $J/\psi$ 
at $Q^2 \simeq 30\gev^2$~\cite{epj:c13:371}.

In QCD-based models, at high $Q^2$, 
the size of the $q\bar{q}$ pair
in the direction transverse to the reaction axis
decreases as $1/Q$ 
and the $t$ dependence should reach a universal limit, 
independent of the flavour of the quark constituents 
of the meson~\cite{pr:d50:3134}. 
Hence, in this limit, the
$t$ dependence is given solely by the GPDs of the nucleon.
Following this idea, the differential cross section was also fitted using 
an elastic form factor for two-gluon exchange,  
$d\sigma /dt \propto (1-t/m^2_{2g})^{-4}$,
where $m^2_{2g}$ is the square of the two-gluon invariant mass,
as suggested by Frankfurt and Strikman \cite{pr:d66:031502}.
The fit, including both statistical and systematic uncertainties,
yields $m^2_{2g}=0.55\pm 0.02\gev^2$ 
and is shown in Fig.~\ref{fig-dsigmadt}a.

\subsection{Pomeron trajectory}
Soft diffractive processes are described by Regge 
phenomenology~\cite{collins:1977:regge}
in terms of the exchange of a Pomeron trajectory.
In hard interactions, where Regge phenomenology may not be applicable,
an effective Pomeron trajectory may nevertheless be extracted.
The high-$Q^2$ sample was analysed to determine the effective 
Pomeron trajectory.
In the Regge formalism, the differential cross section can be expressed as
\begin{equation}
\label{regge1}
d\sigma/dt \propto W^{4(\alpha_{\pom}(t)-1)},
\end{equation}
where the trajectory $\alpha_{\pom}$ is usually parametrised as
\begin{equation*}
\alpha_{\pom}(t)= \alpha_{\pom}(0)+\alpha_{\pom}^\prime t.
\end{equation*}
The effective Pomeron trajectory was determined 
by fitting Eq.~(\ref{regge1}) to the differential cross sections 
at different $t$ values.
The fit was performed in four $t$ bins at $Q^2=6.8\gev^2$. 
Since the proton-dissociative process has the same $W$
dependence as the exclusive process, the extraction of $\alpha_{\pom}$
is not sensitive to this background contribution, which populates
the high-$t$ region. Therefore the analysis was extended up to
$|t|=2\gev^2$.
The fit results are shown in Fig.~\ref{fig-regge} 
and in Table~\ref{tab-regge}. 
The parameters of the trajectory, determined from the linear fit are:
\begin{align*}
& \alpha_{\pom}(0)= 1.20\pm 0.03({\rm stat.})^{+0.01}_{-0.02}({\rm syst.});\\
& \alpha_{\pom}^\prime= 0.07\pm0.05({\rm stat.})^{+0.03}_{-0.04}({\rm syst.})\gev^{-2}.
\end{align*}
These values are in good agreement with the ZEUS results from $J/\psi$
photoproduction~\cite{epj:c24:345} 
which are also shown in Fig.~\ref{fig-regge}. 
They are also in agreement with expectations 
of pQCD-based models~\cite{jetp:70:155,jhep:103:45},
but are not consistent with
the trajectory measured in soft diffractive processes,
$\alpha_{\pom} =1.08 + 0.25 ~t$~\cite{pl:b348:213,pr:d10:170}.  

\subsection{Decay angular distributions}
The study of the angular distributions of the decay of the $J/\psi$
provides information about the photon and $J/\psi$ polarisation states.
In the helicity frame~\cite{np:b61:381},
the production and decay of the $J/\psi$ 
can be described in terms of three angles: 
$\Phi_h$,  the angle between the $J/\psi$ production plane
and the lepton scattering plane; 
$\theta_h$, the polar angle, 
and $\phi_h$, the azimuthal angle of the positively charged decay lepton. 
Under the assumption of SCHC, 
the normalised angular distribution depends only on two angles,
$\theta_h$ and $\psi_h = \phi_h - \Phi_h$,
and can be expressed in the form
\begin{equation}\label{eq-angtheta}
\frac{1}{N}\frac{dN}{d\cos \theta_h} =
\frac{3}{8}  \left[  1 + r^{04}_{00} + (1-3r^{04}_{00})\cos ^2 \theta_h  \right],
\end{equation}
\begin{equation}\label{eq-angpsi}
\frac{1}{N}\frac{dN}{d\psi_h}  =
\frac{1}{2\pi} \left[  1 -  \epsilon r^{1}_{1-1} \cos  2\psi_h  \right].
\end{equation}
The spin-density matrix element $r^{04}_{00}$ represents the probability 
that the  $J/\psi$  is produced in the helicity-0 state 
from a virtual photon of helicity 0 or 1. 
The spin-density matrix element $r^{1}_{1-1}$ gives the probability 
for  the  $J/\psi$ to be  produced in the helicity-1 state 
from a virtual photon of helicity $1$ or $-1$.
Assuming SCHC and natural spin-parity exchange (NPE)~\cite{np:b61:381},
the matrix elements $r^{04}_{00}$ and $r^{1}_{1-1}$ are related  by
\begin{equation}\label{eq-rrel}
r^{1}_{1-1} =  \frac{1}{2} \left (  1-r^{04}_{00} \right ).
\end{equation}

The cross sections at $W=90\gev$ are shown in Fig.~\ref{fig-angular}a-f
for three intervals of $Q^2$.
Equations (\ref{eq-angtheta}) and (\ref{eq-angpsi}) were fitted to the data.
The values of the spin-density matrix elements 
$r^{04}_{00}$ and $r^{1}_{1-1}$,
determined from the fits, are given in Table~\ref{tab-spinmatrix}. 
The measured values of $r^{1}_{1-1}$ are consistent with those
obtained from Eq.~(\ref{eq-rrel}), also shown in Table~\ref{tab-spinmatrix},
supporting the SCHC and NPE hypotheses.

Figures~\ref{fig-RvsW} and \ref{fig-Rvst} show the cross sections
in bins of $W$ and $t$, respectively.
They are quoted at the reference value $Q^2=6.8\gev^2$.
Equation~(\ref{eq-angtheta}) was fitted to the data.
The values of $r^{04}_{00}$, given in Tables~\ref{tab-spinmatrixW}
and \ref{tab-spinmatrixt}, 
are consistent with no $W$ or $t$ dependence.

\subsubsection{Longitudinal and transverse cross sections}
\label{section-R}
The ratio of the longitudinal to transverse cross section,
$R=\sigma_L/\sigma_T$, was calculated as a function of $Q^2$, $W$ and $t$ 
from $r^{04}_{00}$ according to the relation
\begin{equation*}\label{eq-R}
R = \frac{1}{\epsilon}  \frac{r^{04}_{00}}{1-r^{04}_{00}},
\end{equation*}
which is valid under the assumption of SCHC.

The values of $R$ as a function of $Q^2$ are given in 
Table~\ref{tab-spinmatrix} 
and compared with the H1 results~\cite{epj:c10:373}
in Fig.~\ref{fig-angular}g.
The expression $R=\zeta(Q^2/M_{J/\psi}^2$) was fitted to the ZEUS data
yielding $\zeta=0.52\pm0.16({\rm stat.})$.
In QCD-based models, the scale that controls
the transverse size of the $q\bar{q}$ fluctuation of the photon
may behave differently for $\sigma_L$ and $\sigma_T$.
However, in the MRT model,  
$\sigma_L$ and $\sigma_T$ have the same $W$ dependence,
dictated by the gluon distribution. Therefore the ratio is constant.
This model correctly describes the rising behaviour of $R$ with $Q^2$
whereas the GLLMN prediction somewhat overestimates it.

The values of $R$ as a function of $W$ and $t$
are given in Tables~\ref{tab-spinmatrixW} and \ref{tab-spinmatrixt}
and shown in Figs.~\ref{fig-RvsW}f and \ref{fig-Rvst}g, respectively.

\section{Summary}\label{sec-conclusions}
The exclusive electroproduction of $J/\psi$ mesons, 
$ep\rightarrow e J/\psi\:p$, 
has been measured with the ZEUS detector at HERA 
for photon virtualities in the ranges 
$0.15<Q^2<0.8\gev^2$ and $2<Q^2<100\gev^2$, 
for photon-proton centre-of-mass energies in the range $30<W<220\gev$ 
and for four-momentum-transfer squared in the range $|t|<1\gev^2$. 

The cross section of the process $\gamma^*\:p\rightarrow J/\psi\:p$ 
rises with $W$ as $\sigma \propto W^{\delta}$, 
with a slope parameter $\delta$ of about 0.7. 
This parameter does not change significantly with  $Q^2$ and 
is consistent with that observed in $J/\psi$ photoproduction.  

The cross section 
at $W=90\gev$ and over the whole $Q^2$ range
is described by the function
$\sigma\propto (Q^2+M^2_{J/\psi})^{-n}$, with $n=2.44\pm0.08$.   

The $t$ distribution, measured for $|t|<1\gev^2$, is well described by an 
exponential dependence over the range $2<Q^2<100\gev^2$. 
The slope parameter, $b$, is consistent with being constant in this range. 
The mean value is  
$b=4.72\pm0.15({\rm stat.}) \pm 0.12 ({\rm syst.})\gev^{-2}$, 
consistent with that observed in $J/\psi$ photoproduction. 

An analysis of the cross sections in the framework of Regge phenomenology
yields an effective Pomeron trajectory consistent with that measured
in $J/\psi$ photoproduction.

The spin-density matrix elements $r^{1}_{1-1}$ and $r^{04}_{00}$ 
are consistent with $s$-channel-helicity conservation. 
The ratio of the cross sections for longitudinally and transversely
polarised photons, $R$, increases with $Q^2$, 
but is independent of $W$ and $t$, within the measured range.

The $J/\psi$ electroproduction data can be qualitatively described within  
the framework of pQCD
that successfully describes $J/\psi$ photoproduction data.
The data exhibit a strong sensitivity to the gluon distribution in the proton.
Full next-to-leading-order QCD calculations would allow 
these data to be used in global QCD fits to constrain the gluon density
function in the proton.

\section*{Acknowledgements}
We thank the DESY directorate for their strong support and encouragement.
The special effort of the HERA group is gratefully acknowledged.
We are grateful for the support of the DESY computing and network services.
The design, construction and installation of the ZEUS detector 
has been made possible by the efforts and ingenuity of many people 
who are not listed as authors. 
It is a pleasure to thank E.~Naftali and T.~Teubner 
for  providing us with model predictions.

%% file: zdraft-bbl.tex
\providecommand{\etal}{et al.\xspace}
\providecommand{\coll}{Coll.\xspace}
\catcode`\@=11
\def\@bibitem#1{%
\ifmc@bstsupport
  \mc@iftail{#1}%
    {;\newline\ignorespaces}%
    {\ifmc@first\else.\fi\orig@bibitem{#1}}
  \mc@firstfalse
\else
  \mc@iftail{#1}%
    {\ignorespaces}%
    {\orig@bibitem{#1}}%
\fi}%
\catcode`\@=12

%% file: zdraft-tab.tex

\begin{table}[p]
\begin{center}
\footnotesize{%
\begin{tabular}{|c|c|c|c|c|c|c|c|c|c|}
\hline
$Q^2$ &$\langle Q^2 \rangle$& W & $\langle W \rangle$ & $N_{ee}$ & $N_{\mu\mu}$ & $\mathcal{A}_{ee}$&
$\mathcal{A}_{\mu\mu}$& $\sigma^{e p \rightarrow J/\psi p}$ & $\sigma_{tot}^{\gamma^{*} p \rightarrow J/\psi p}$ \\
(GeV$^2$) &(GeV$^2$)&(GeV)&(GeV)& & &&& (pb) & (nb) \\
\hline
\hline
           &     & 30 - 65 &  49 & 32 & & 0.031 & &
$217 \pm 53 ^{+ 12}_{- 19}$ & $ 39.2\pm  9.6^{+ 2.2}_{- 3.4}$\\
0.15 - 0.8 & 0.4 & 65 - 105 &  86 & 55 & & 0.044 & &
$257 \pm 46 ^{+ 18}_{- 17}$ & $ 75.7\pm 13.5^{+ 5.2}_{- 4.9}$\\
           &     &105 - 220 & 167 & 50 & & 0.021 & &
$498 \pm 89 ^{+ 37}_{- 38}$ & $118.0\pm 21.0^{+ 8.8}_{- 9.1}$\\

\hline
      & &30 - 45   & 37  & 29.2 &     & 0.111&      & $41.5  \pm 8.4  ^{+ 5.6 }_{- 6.6}$& $24.8\pm5.0^{+3.3}_{-3.9}$\\
      & &45 - 70   & 57  & 51.5 & 53.2& 0.180& 0.173& $48.8  \pm 5.2  ^{+ 3.1 }_{- 3.9}$& $27.4\pm3.0^{+1.8}_{-2.2}$\\
2 - 5 &3.1&70 - 90   & 80  & 36.7 & 60.0& 0.204& 0.224&$36.4  \pm 4.1  ^{+10.5 }_{- 3.0}$& $36.7\pm4.2^{+10.6}_{-3.0}$\\
      & &90 - 112  & 101 & 61.7 & 37.5& 0.221& 0.223& $35.4  \pm 4.0  ^{+ 3.0 }_{- 4.5}$& $43.0\pm4.8^{+3.7}_{-5.4}$\\
      & &112 - 145 & 128 & 51.2 & 46.4& 0.197& 0.167& $44.7  \pm 5.0  ^{+ 9.0 }_{- 4.3}$& $48.8\pm5.5^{+9.8}_{-4.7}$\\
      & &145 - 220 & 180 & 71.6 &     & 0.154&      & $76.5  \pm10.3  ^{+11.5 }_{- 5.1}$& $61.1\pm8.2^{+9.2}_{-4.1}$\\
\hline
      & &30 - 50  &  40 & 27.8 &     & 0.215&      & $19.6  \pm 4.1  ^{+ 3.9 }_{- 1.9}$& $12.7\pm2.7^{+2.5}_{-1.2}$\\
      & &50 - 74  &  62 & 48.7 & 45.8& 0.403& 0.383& $19.3  \pm 2.2  ^{+ 2.9 }_{- 1.3}$& $16.6\pm1.9^{+2.5}_{-1.1}$\\
5 - 10& 6.8&74 - 96  &  85 & 39.6 & 52.4& 0.439& 0.486& $15.6  \pm 1.8  ^{+ 1.6 }_{- 1.4}$& $20.7\pm2.3^{+2.2}_{-1.9}$\\
      & &96 - 120 & 108 & 37.1 & 46.4& 0.479& 0.514& $13.5  \pm 1.7  ^{+ 1.1 }_{- 0.7}$& $21.9\pm2.7^{+1.7}_{-1.2}$\\
      & &120 - 150& 135 & 33.9 & 49.0& 0.475& 0.395& $14.9  \pm 1.9  ^{+ 1.1 }_{- 1.3}$& $25.8\pm3.3^{+1.9}_{-2.3}$\\
      & &150 - 220& 183 & 58.4 &     & 0.343&      & $27.9  \pm 4.1  ^{+ 4.5 }_{- 1.4}$& $33.2\pm4.9^{+5.3}_{-1.6}$\\
\hline
         & &30 - 55  &  42 & 16.1&     & 0.235&      & $10.9  \pm 3.1  ^{+ 0.8 }_{- 1.0}$& $3.3\pm0.9^{+0.2}_{-0.3}$\\
         & &55 - 78  &  66 & 27.7& 37.8& 0.555& 0.659& $ 8.4  \pm 1.2  ^{+ 1.4 }_{- 0.4}$& $4.5\pm0.6^{+0.7}_{-0.2}$\\
10 - 100 & 16.0&78 - 100 &  89 & 31.0& 43.6& 0.673& 0.728& $ 8.6  \pm 1.1  ^{+ 0.9 }_{- 1.4}$& $6.7\pm0.9^{+0.7}_{-1.1}$\\
         & &100 - 124& 112 & 37.5& 36.2& 0.645& 0.704& $ 8.4  \pm 1.1  ^{+ 0.4 }_{- 1.2}$& $7.9\pm1.0^{+0.3}_{-1.1}$\\
         & &124 - 160& 141 & 39.6& 41.3& 0.563& 0.591& $10.8  \pm 1.4  ^{+ 2.1 }_{- 0.8}$& $9.3\pm1.2^{+1.7}_{-0.7}$\\
         & &160 - 220& 189 & 51.0&     & 0.361&      & $25.1  \pm 3.8  ^{+ 1.7 }_{- 1.2}$& $20.8 \pm3.1^{+1.4}_{-1.0}$\\
\hline
\end{tabular}}
\caption{
The cross sections for the reaction $\gamma^* p \rightarrow J/\psi \: p$ 
measured as a function of $W$ in bins of $Q^2$ and for $|t|<1\gev^2$:
$\langle W \rangle$ and $\langle Q^2 \rangle$ 
are the mean values in the indicated ranges;
$N_{ee}$ and $N_{\mu\mu}$ are the number of events in the signal region 
after non-resonant background subtraction 
of the electron and muon pairs, respectively;
${\cal A}_{ee}$ and ${\cal A}_{\mu\mu}$ are the corresponding acceptances.
The first uncertainty of the cross sections is statistical 
and the second systematic.
An overall normalisation uncertainty of $^{+5\%}_{-8\%}$ was not included.
}
\label{tab-Wdep}
\end{center}
\end{table}

\begin{table}[p]
\begin{sideways}\begin{minipage}[b]
{\textheight}
\vspace{2cm}    
\begin{center}
\begin{tabular}{|c|c|c|c|c|c|c|c|c|c|}
\hline
$Q^2$ & $\langle Q^2 \rangle$ & 
$N_{ee}$ & $\mathcal{A}_{ee}$& $\sigma^{e p \rightarrow J/\psi p}_{ee}$ & 
$N_{\mu\mu}$ & $\mathcal{A}_{\mu\mu}$ &$\sigma^{e p \rightarrow J/\psi p}_{\mu\mu}$ & 
$\sigma_{tot}^{\gamma^{*} p \rightarrow J/\psi p}$ \\ 
(GeV$^2$) &(GeV$^2$)& & &(pb)& & &(pb)&(nb)\\ 
\hline

\multicolumn{2}{|c|}{}& 
\multicolumn{3}{c|}{$ee$: $30<W<220\gev$}&
\multicolumn{3}{c|}{$\mu\mu$: $45<W<160\gev$}&
$W=90\gev$\\
\hline
\hline
0.15 - 0.8 & 0.4 & 137.0 & 0.029 & $954 \pm108 ^{+ 63}_{- 74}$ 
& & & & $ 72.6 \pm  8.2^{+ 4.8}_{- 5.6}$\\

\hline
 2   -  3.2& 2.5 & 141.3 & 0.156 & $ 150 \pm 14 ^{+  53}_{- 8}$ & 90.5 & 0.159 & $96  \pm11  ^{+ 5 }_{-14}$& $39.7 \pm 2.9^{+5.9}_{-2.9}$\\
 3.2 -  5  & 3.9 & 160.6 & 0.202 & $ 132 \pm 12 ^{+ 8}_{- 17}$& 118.6 & 0.217 &$91.6  \pm 8.9  ^{+12.2 }_{- 6.6}$& $38.7 \pm 2.5^{+3.3}_{-3.6}$\\
 5   -  7  & 5.9 & 123.1 & 0.327 & $59.9 \pm 6.1 ^{+ 5.5 }_{- 3.6}$ & 100.4 & 0.336 &$48.7  \pm 5.2  ^{+ 1.2 }_{- 2.5}$& $24.3 \pm 1.8^{+1.1}_{-1.0}$\\
 7   - 10  &  8.4 & 122.5 & 0.466 & $42.6 \pm 4.3 ^{+ 4.7 }_{- 5.0}$ & 104.1 & 0.529 & $32.5  \pm 3.4  ^{+ 4.4 }_{- 2.4}$& $15.8 \pm 1.2^{+1.5}_{-0.7}$\\
10   - 15  & 12  & 106.7 & 0.485 & $36.7 \pm 4.0 ^{+ 1.4 }_{- 2.9}$  &  87.2 & 0.607  & $24.1  \pm 2.8  ^{+ 1.2 }_{- 1.6}$& $11.6 \pm 1.0^{+0.4}_{-0.6}$\\
15   - 40  & 22  &  84.3 & 0.473 & $29.3 \pm 3.7 ^{+ 2.0 }_{- 4.7}$  &  71.6 & 0.651 &$18.4  \pm 2.4  ^{+ 0.9 }_{- 1.4}$& $ 4.0 \pm 0.4^{+0.2}_{-0.3}$\\
40   - 100 & 54  &  12.0 & 0.423 & $ 4.5\pm 1.5^{+0.5}_{-1.1}$  &   7.4 & 0.554 &$ 2.2\pm 0.9^{+0.4}_{-0.6}$& $ 0.65\pm 0.17^{+0.08}_{-0.16}$\\
\hline
\end{tabular}
\hspace*{-5.5cm}\begin{minipage}[b]{27.5cm}
\setlength{\captionwidth}{20.0cm}
\isucaption{
The cross sections for the reaction $\gamma^* p \rightarrow J/\psi \: p$ 
measured as a function of $Q^2$,
for a mean value $W=90\gev$ and for $|t|<1\gev^2$:
$\langle Q^2 \rangle$ indicates the mean value in the $Q^2$ range considered;
$N_{ee}$ and $N_{\mu\mu}$ are the numbers of events in the signal region 
after non-resonant background subtraction
of the electron and muon pairs, respectively;
${\cal A}_{ee}$ and ${\cal A}_{\mu\mu}$ are the corresponding acceptances.
The first uncertainty of the cross sections is statistical 
and the second systematic.
An overall normalisation uncertainty of $^{+5\%}_{-8\%}$ was not included.
}
\label{tab-Q2dep}
\end{minipage}
\end{center}
\end{minipage}\end{sideways}
\end{table}

\begin{table}[p]
\begin{center}
\footnotesize{%
\begin{tabular}{|c|c|c|c|c|}
\hline
$Q^2$ &$\langle Q^2 \rangle$& $|t|$ & $\langle |t| \rangle$ & 
$d\sigma^{\gamma^* p \rightarrow J/\psi p} / dt$ \\
(GeV$^2$) &(GeV$^2$)&(GeV$^2$)&(GeV$^2$) & (nb/GeV$^2$) \\
\hline
\hline
       &    & 0.0 - 0.1 & 0.05 & $79.2  \pm 5.0  ^{+ 6.1 }_{- 6.5}$\\
2 - 100& 6.8& 0.1 - 0.2 & 0.15 & $43.9  \pm 3.0  ^{+ 2.8 }_{- 1.9}$\\
       &    & 0.2 - 0.4 & 0.29 & $25.8  \pm 1.4  ^{+ 1.9 }_{- 1.5}$\\
       &    & 0.4 - 1.0 & 0.58 & $ 6.0  \pm 0.4  ^{+ 0.4 }_{- 0.4}$\\
\hline
      &    & 0.0 - 0.1 & 0.05 & $148  \pm 15  ^{+ 22 }_{- 14}$\\
2 - 5 & 3.1& 0.1 - 0.2 & 0.15 & $86.9  \pm 9.6  ^{+12.5 }_{- 8.1}$\\
      &    & 0.2 - 0.4 & 0.29 & $49.2  \pm 4.3  ^{+ 6.6 }_{- 2.8}$\\
      &    & 0.4 - 1.0 & 0.58 & $10.7  \pm 1.1  ^{+ 0.9 }_{- 0.7}$\\
\hline
       &     & 0.0 - 0.1 & 0.05 & $75.6  \pm 8.3  ^{+ 5.0 }_{- 9.7}$\\
5 - 10 & 6.8 & 0.1 - 0.2 & 0.15 & $39.6  \pm 4.9  ^{+ 2.0 }_{- 2.4}$\\
       &     & 0.2 - 0.4 & 0.29 & $23.9  \pm 2.4  ^{+ 2.1 }_{- 1.1}$\\
       &     & 0.4 - 1.0 & 0.58 & $ 6.5  \pm 0.6  ^{+ 0.6 }_{- 0.3}$\\
\hline
       &   & 0.0 - 0.1 & 0.05 & $28.0  \pm 3.4  ^{+ 2.6 }_{- 2.8}$\\
10-100 & 16& 0.1 - 0.2 & 0.15 & $15.6  \pm 2.0  ^{+ 0.9 }_{- 2.0}$\\
       &   & 0.2 - 0.4 & 0.29 & $ 9.4  \pm 1.0  ^{+ 0.6 }_{- 1.4}$\\
       &   & 0.4 - 1.0 & 0.58 & $ 1.8  \pm 0.2  ^{+ 0.1 }_{- 0.2}$\\
\hline
\end{tabular}}
\caption{
The differential cross sections for the reaction $\gamma^* p \rightarrow J/\psi \: p$ 
measured as a function of $t$ in bins of $Q^2$ for a mean value $W=90\gev$. 
The first uncertainty is statistical and the second systematic.
}
\label{tab-tdep}
\end{center}
\end{table}

\begin{table}
\begin{center}
\begin{tabular}{|c|c|c|c|c|}
\hline
$Q^2$  & $\langle Q^2 \rangle$  &$\delta $&$b$ $(\frac{d\sigma}{dt} \propto e^{-b|t|})$ & $\left.\frac{d\sigma}{dt}\right|_{t=0}$\\ 
 (GeV$^2$) & (GeV$^2$) &$(\sigma\propto W^{\delta})$& (GeV$^{-2}$)& (nb/GeV$^2$) \\ 
\hline
\hline
0.15 - 0.8& 0.4 & $ 0.87\pm 0.22 ^{+0.04}_{-0.01} $ &&\\
\hline
2 - 5  & 3.1& $0.65 \pm 0.17^{+0.16}_{-0.05} $ &$ 4.85\pm 0.24 ^{+0.26}_{-0.19}$  & $ 185 \pm 15 ^{+30}_{-21}$\\
5 - 10 & 6.8& $0.60 \pm 0.18^{+0.04}_{-0.10} $ &$ 4.44\pm 0.26 ^{+0.12}_{-0.27}$  & $  84.7 \pm  7.9 ^{+7.3}_{-9.6}$\\
10 -100&  16& $1.12 \pm 0.20^{+0.03}_{-0.16} $ &$ 5.06\pm 0.27 ^{+0.29}_{-0.17}$  & $  35.5 \pm  3.4 ^{+2.9}_{-3.5}$\\
\hline
2 -100& 6.8& $0.73\pm0.11^{+0.04}_{-0.08} $ &$4.72\pm 0.15 ^{+0.12}_{-0.12}$&$ 95.2 \pm 4.9 ^{+8.1}_{-7.9} $\\
\hline
\end{tabular}
\caption{
The parameters $\delta$, $b$ and $\frac{d\sigma}{dt}|_{t=0}$ 
measured as a function of $Q^2$ in the range $30<W<220\gev$ and $45<W<160\gev$ for the electron and muon channels, respectively,  
and $|t|<1\Gev^2$.
The first uncertainty is statistical and the second systematic.
}
\label{tab-slopes}
\end{center}
\end{table}

\begin{table}
\begin{center}
\begin{tabular}{|c|c|c|}
\hline
$|t|$ & $\langle |t| \rangle$  &$\alpha_{I\!P}(t)$\\ 
 (GeV$^2$) & (GeV$^2$) &\\ 
\hline
\hline
0.0 - 0.1& 0.046 & $1.22\pm 0.04 ^{+0.03}_{-0.04}$\\
0.1 - 0.3& 0.186 & $1.17\pm 0.04 ^{+0.02}_{-0.02}$\\
0.3 - 0.9& 0.483 & $1.17\pm 0.03 ^{+0.02}_{-0.04}$\\
0.9 - 2.0& 1.123 & $1.13\pm 0.04 ^{+0.03}_{-0.04}$\\
\hline
\end{tabular}
\caption{
The Pomeron trajectory $\alpha_{\pom}(t)$ 
measured in four $t$ bins, in the range $2<Q^2<100\gev^2$
at a mean value $\langle Q^2 \rangle =6.8\gev^2$.
The first uncertainty is statistical and the second systematic.
}
\label{tab-regge} 
\end{center}
\end{table}

\begin{table}
\begin{center}
\begin{tabular}{|c|c|c|c|c|c|}
\hline
$Q^2$  & $\langle Q^2 \rangle$  &$r^{04}_{00}$&$r^{1}_{1-1}$&$R=\sigma_L/\sigma_T$& $r^{1}_{1-1} -  \frac{1}{2} \left (  1-r^{04}_{00} \right )$\\ 
 (GeV$^2$) & (GeV$^2$) &&&& \\ 
\hline
\hline
2 - 5   & 3.1 &$0.12\pm 0.08^{+0.13}_{-0.15}$ & $0.34\pm 0.09^{+0.03}_{-0.06}$ & $0.13\pm 0.11^{+0.09}_{-0.16}$ & $-0.10\pm 0.09^{+0.08}_{-0.06}$\\
5 - 10  & 6.8 &$0.25\pm 0.09^{+0.10}_{-0.06}$ & $0.44\pm 0.09^{+0.06}_{-0.07}$ & $0.33\pm 0.16^{+0.19}_{-0.11}$ & $ 0.06\pm 0.10^{+0.08}_{-0.06}$\\
10 -100 & 16  &$0.54\pm 0.10^{+0.06}_{-0.03}$ & $0.26\pm 0.09^{+0.09}_{-0.04}$ & $1.19\pm 0.51^{+0.28}_{-0.14}$ & $ 0.03\pm 0.11^{+0.07}_{-0.02}$\\
\hline
\end{tabular}
\caption{The spin-density matrix elements, $r^{04}_{00}$ and $r^{1}_{1-1}$, 
the ratio of cross sections of longitudinally and transversely 
polarised photons, $R$, and the quantity $r^{1}_{1-1} -  \frac{1}{2} \left (  1-r^{04}_{00} \right )$ measured in bins of $Q^2$.
The first uncertainty is statistical and the second systematic.
}
\label{tab-spinmatrix} 
\end{center}
\end{table}

\begin{table}
\begin{center}
\begin{tabular}{|c|c|c|c|}
\hline
$W$  & $\langle W \rangle$  &$r^{04}_{00}$&$R=\sigma_L/\sigma_T$\\
 (GeV) & (GeV) &&\\
\hline
\hline
  30 -  55 & 43.5 & 0.21$\pm 0.16^{+0.32}_{-0.18}$ & 0.27$\pm 0.26^{+0.45}_{-0.17}$ \\
  55 -  80 & 68.1 & 0.24$\pm 0.13^{+0.10}_{-0.10}$ & 0.31$\pm 0.23^{+0.26}_{-0.22}$ \\
  80 - 120 & 95.6 & 0.25$\pm 0.09^{+0.09}_{-0.05}$ & 0.33$\pm 0.16^{+0.15}_{-0.07}$ \\
 120 - 160 & 128.1 & 0.12$\pm 0.11^{+0.11}_{-0.05}$ & 0.14$\pm 0.15^{+0.12}_{-0.05}$ \\
 160 - 220 & 184.4 & 0.36$\pm 0.16^{+0.12}_{-0.10}$ & 0.56$\pm 0.40^{+0.23}_{-0.16}$ \\
\hline
\end{tabular}
\caption{The spin density matrix element $r^{04}_{00}$ and 
the ratio of cross sections of longitudinally and transversely polarised 
photons, $R$,
measured in bins of $W$, in the range $2 < Q^2 < 100 \gev^2$ 
at a mean value $\langle Q^2 \rangle=6.8\gev^2$.
The first uncertainty is statistical and the second systematic.
}
\label{tab-spinmatrixW}
\end{center}
\end{table}

\begin{table}
\begin{center}
\begin{tabular}{|c|c|c|c|}
\hline
$|t|$  & $\langle |t| \rangle$  &$r^{04}_{00}$&$R=\sigma_L/\sigma_T$\\
 (GeV$^2$) & (GeV$^2$) &&\\
\hline
\hline
  0.0 - 0.1 & 0.046 & 0.24$\pm 0.11^{+0.12}_{-0.06}$ & 0.31$\pm 0.19^{+0.22}_{-0.10}$ \\
  0.1 - 0.2 & 0.146 & 0.36$\pm 0.13^{+0.08}_{-0.11}$ & 0.56$\pm 0.30^{+0.17}_{-0.20}$ \\
  0.2 - 0.4 & 0.285 & 0.19$\pm 0.10^{+0.07}_{-0.12}$ & 0.23$\pm 0.15^{+0.11}_{-0.16}$ \\
  0.4 - 1.0 & 0.579 & 0.16$\pm 0.10^{+0.05}_{-0.05}$ & 0.19$\pm 0.14^{+0.08}_{-0.08}$ \\
\hline
\end{tabular}
\caption{The spin density matrix element $r^{04}_{00}$ and 
the ratio of cross sections of longitudinally and transversely polarised 
photons, $R$,
measured in bins of $|t|$, in the range $2 < Q^2 < 100 \gev^2$ 
at a mean value $\langle Q^2 \rangle=6.8\gev^2$.
The first uncertainty is statistical and the second systematic.
}
\label{tab-spinmatrixt}
\end{center}
\end{table}

%% file: zdraft-fig.tex
\begin{figure}[p]
\vfill
\begin{center}
\epsfig{file=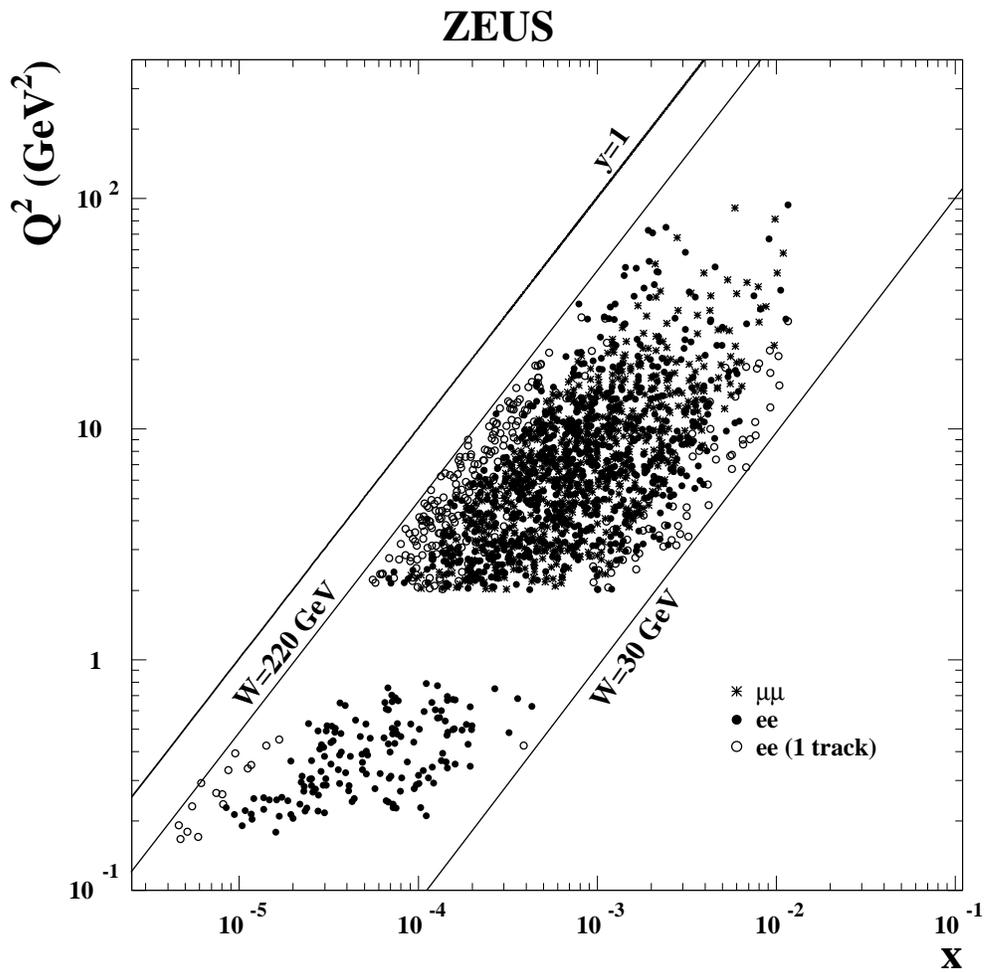,width=13cm}
\end{center}
\caption{The distribution of the events in the muon and electron channels  
in the kinematic plane 
of Bjorken-x and $Q^2$.
The events reconstructed using one and two
measured tracks are shown separately.}
\label{fig-scattq2x}
\vfill
\end{figure} 

\begin{figure}[p]
\vfill
\begin{center}
\epsfig{file=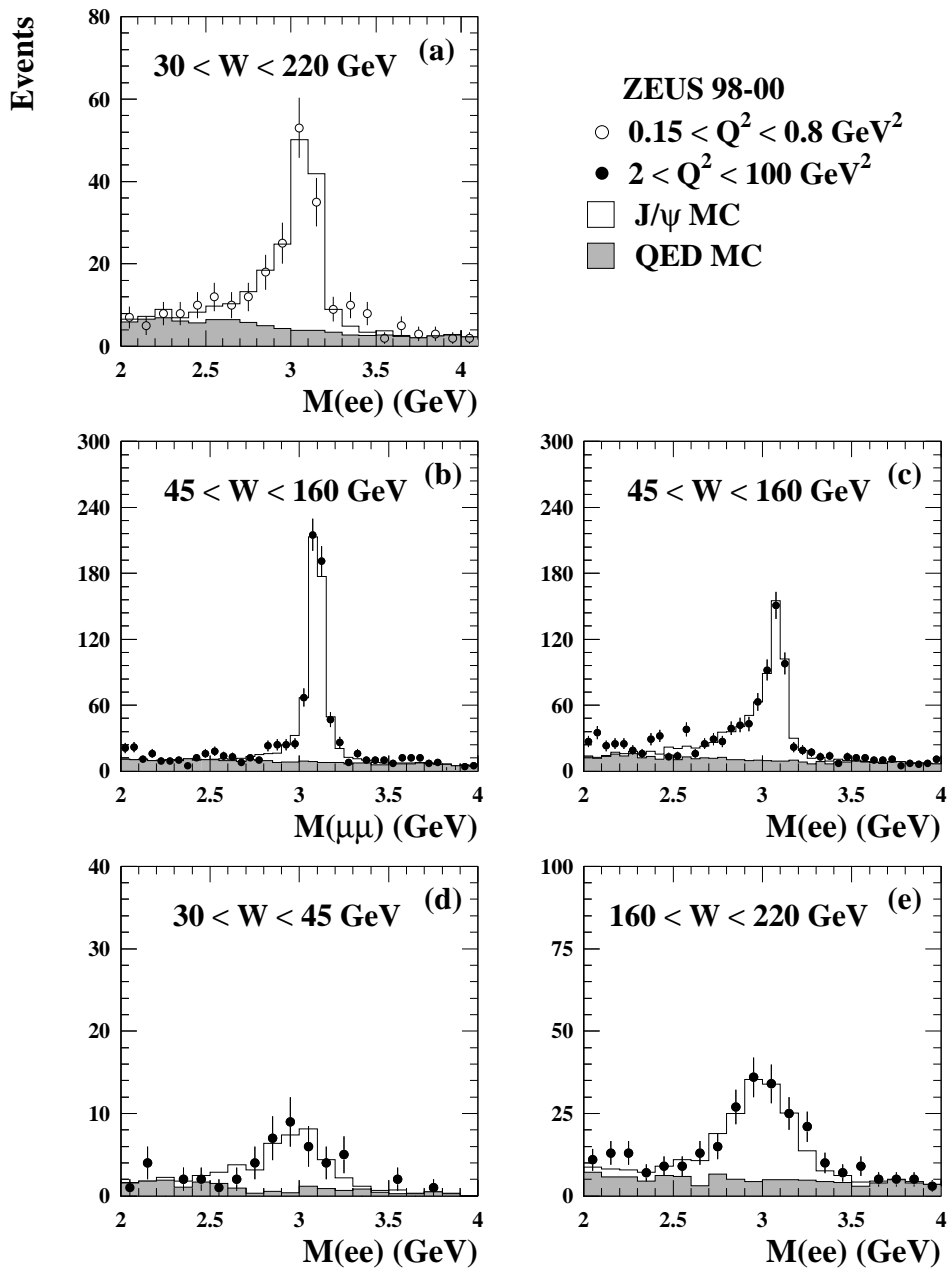,width=13cm}
\end{center}
\caption{
Invariant mass distributions of the lepton pairs for 
(a) the low-$Q^2$  sample and (b)-(e) the high-$Q^2$ sample. 
The shaded histograms are the QED MC distributions
and the open histograms the sum of the $J/\psi$ and QED MC events.
The small excess of data at low mass is due to background from pions.
The error bars indicate the statistical uncertainties.
}
\label{fig-inv_mass}
\vfill
\end{figure}

\begin{figure}[p]
\vfill
\begin{center}
\epsfig{file=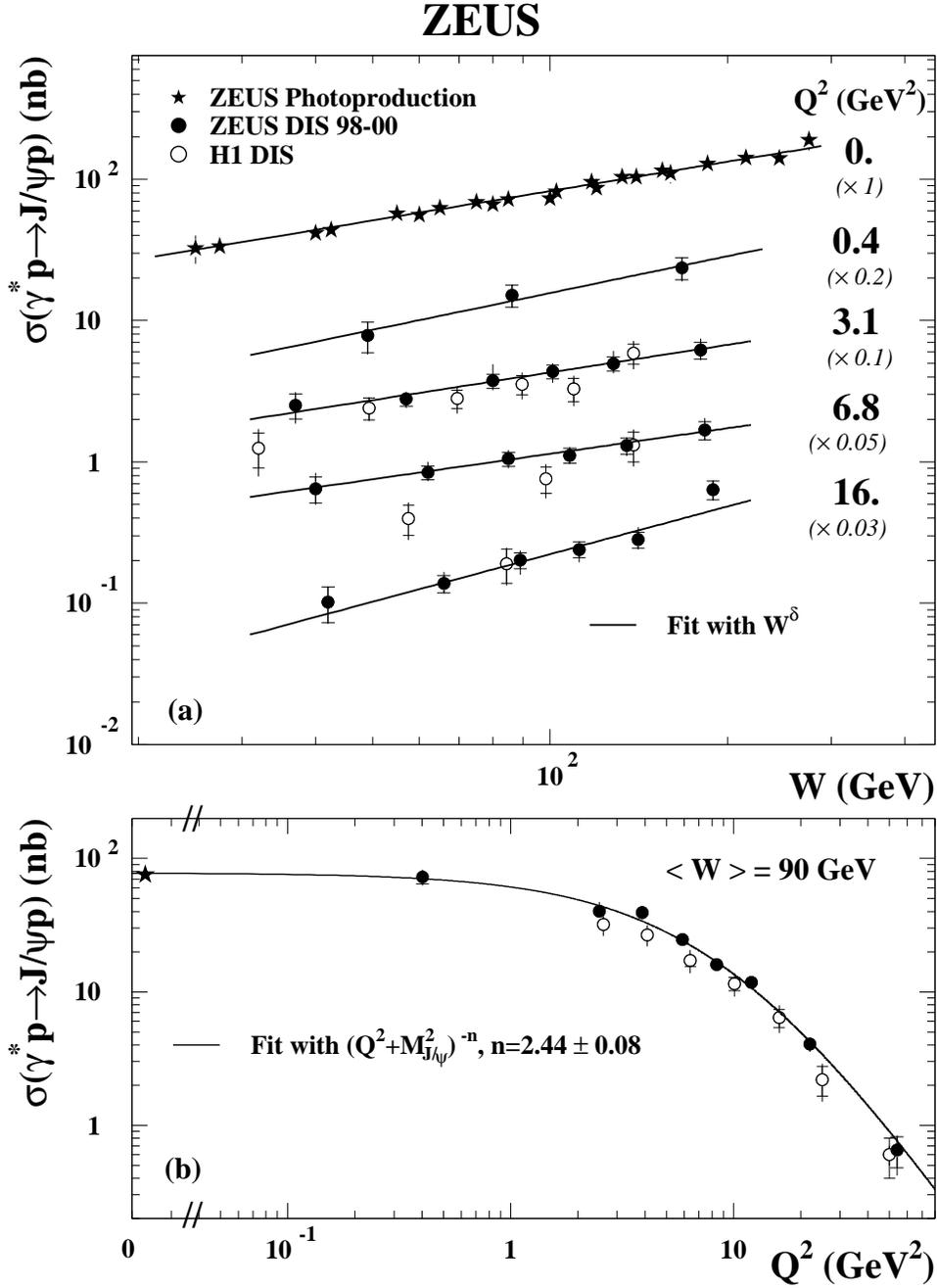,width=13cm}
\end{center}
\caption{
Exclusive $J/\psi$ electroproduction cross section 
(a) as a function of $W$ for four values of $Q^2$
and (b) as a function of $Q^2$ at $\langle W\rangle=90\gev$.
ZEUS photoproduction and H1 electroproduction cross sections 
are also shown.
The full lines are fits to the ZEUS data  
of the form (a) $\sigma \propto W^{\delta(Q^2)}$ and
(b) $\sigma \propto (Q^2+M_{J/\psi}^2)^{-n}$.
The inner error bars represent the statistical uncertainties,
the outer bars are the statistical and systematic uncertainties added in
quadrature.
An overall normalisation uncertainty of $^{+5\%}_{-8\%}$ was not included.
}
\label{fig-xsec}
\vfill
\end{figure}

\begin{figure}[p]
\vfill
\begin{center}
\epsfig{file=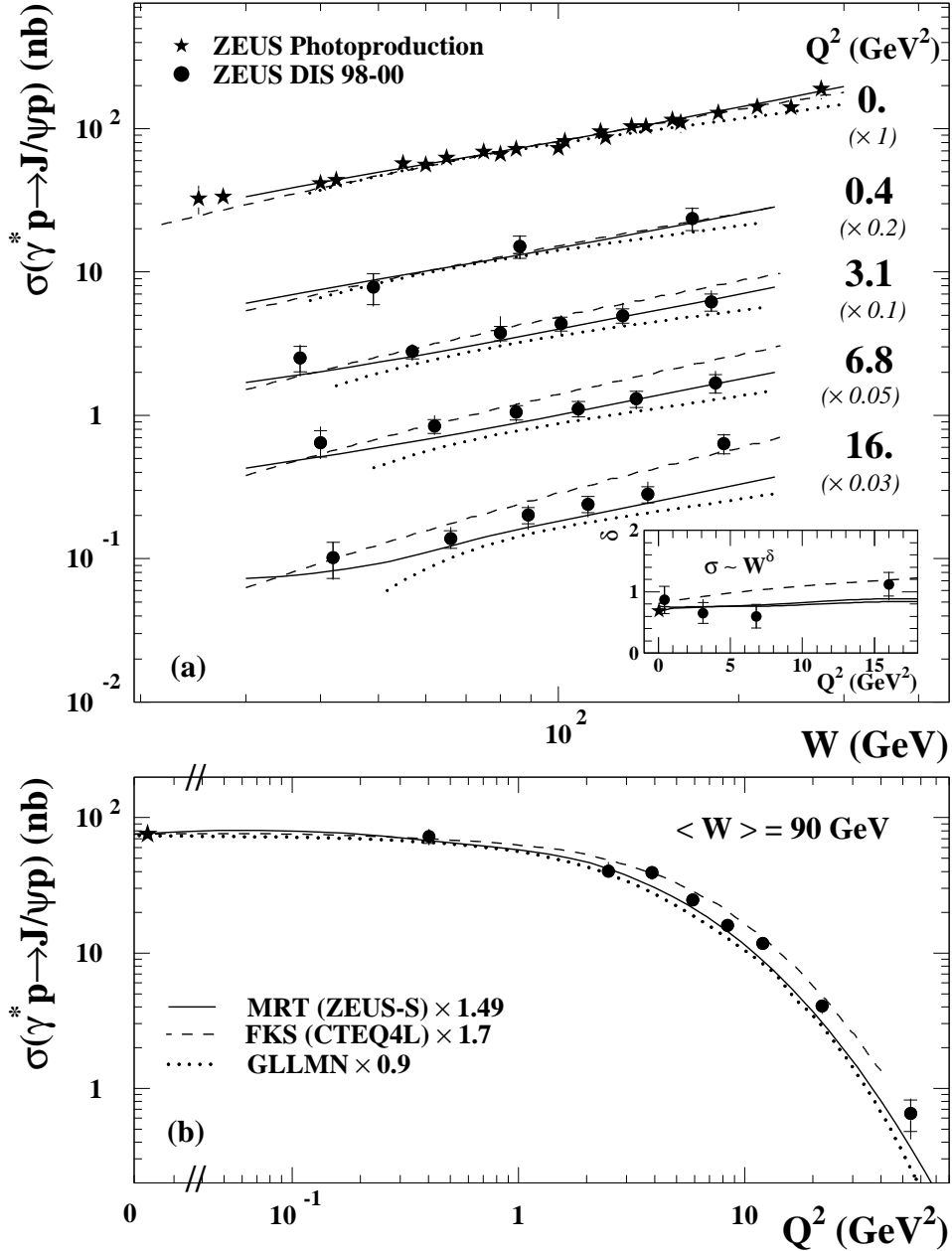,width=13cm}
\end{center}
\caption{
Exclusive $J/\psi$ electroproduction cross section 
(a) as a function of $W$ for four values of $Q^2$
and (b) as a function of $Q^2$ at $\langle W\rangle=90\gev$.
ZEUS photoproduction results are also shown.
The curves represent the predictions of the QCD models MRT,
FKS and GLLMN (see text) normalised to the ZEUS photoproduction point at $\langle W\rangle=90\gev$. 
The insert shows the parameter $\delta$ as a function of $Q^2$.
The inner error bars represent the statistical uncertainties,
the outer bars are the statistical and systematic uncertainties added in
quadrature. 
An overall normalisation uncertainty of $^{+5\%}_{-8\%}$ was not included.
}
\label{fig-xsec-models}
\vfill
\end{figure}

\begin{figure}[p]
\vfill
\begin{center}
\epsfig{file=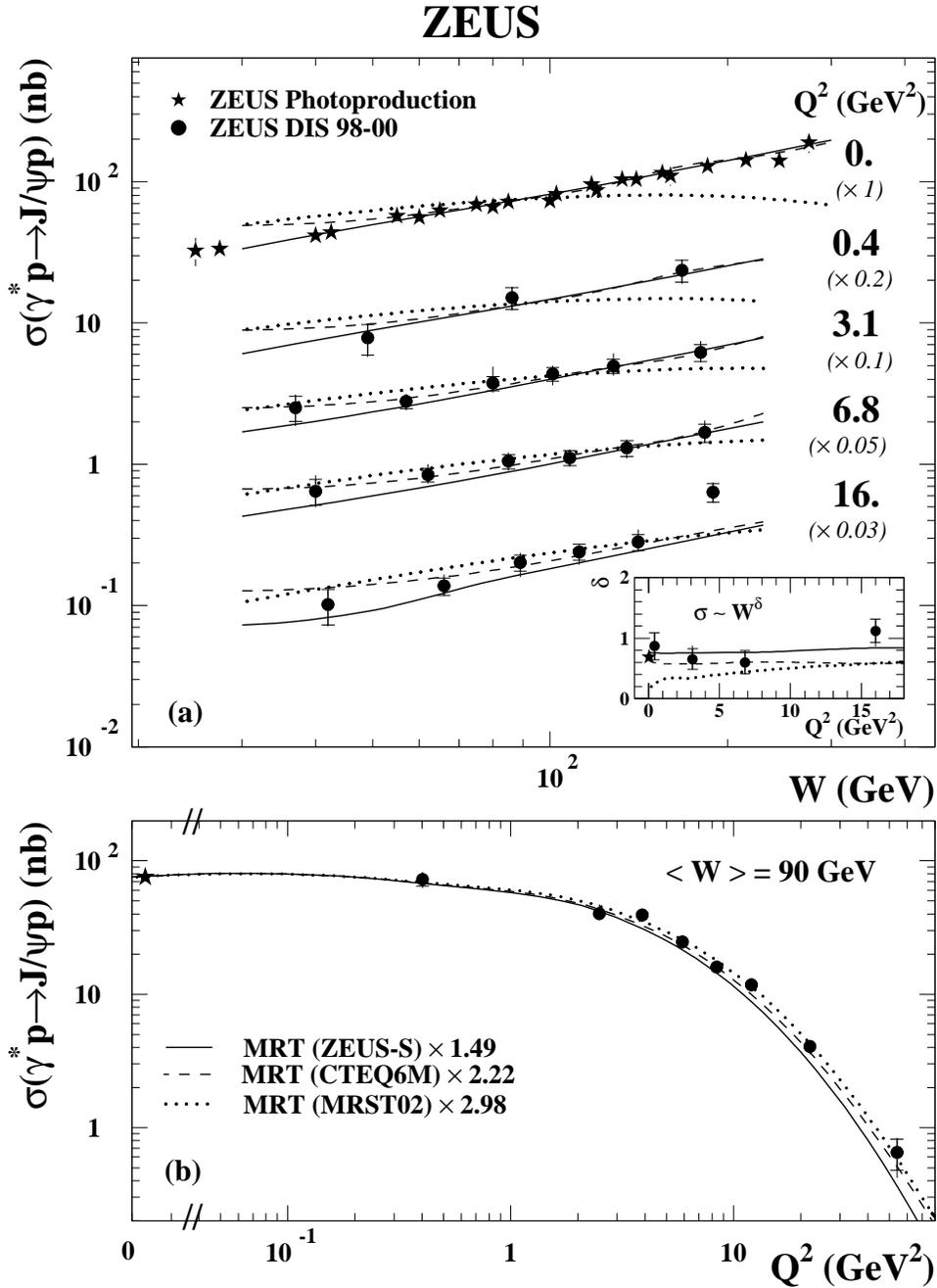,width=13cm}
\end{center}
\caption{
Exclusive $J/\psi$ electroproduction cross section 
(a) as a function of $W$ for four values of $Q^2$
and (b) as a function of $Q^2$ at $\langle W\rangle=90\gev$.
ZEUS photoproduction results are also shown.
The data are compared to the MRT predictions (see text) 
obtained with different parametrisations of the gluon density
and normalised to the ZEUS photoproduction point at $\langle W\rangle=90\gev$. 
The insert shows the parameter $\delta$ as a function of $Q^2$.
The inner error bars represent the statistical uncertainties,
the outer bars are the statistical and systematic uncertainties added in
quadrature. 
An overall normalisation uncertainty of $^{+5\%}_{-8\%}$ was not included.
}  
\label{fig-xsec_gluons}
\vfill
\end{figure}

\begin{figure}[p]
\vfill
\begin{center}
\epsfig{file=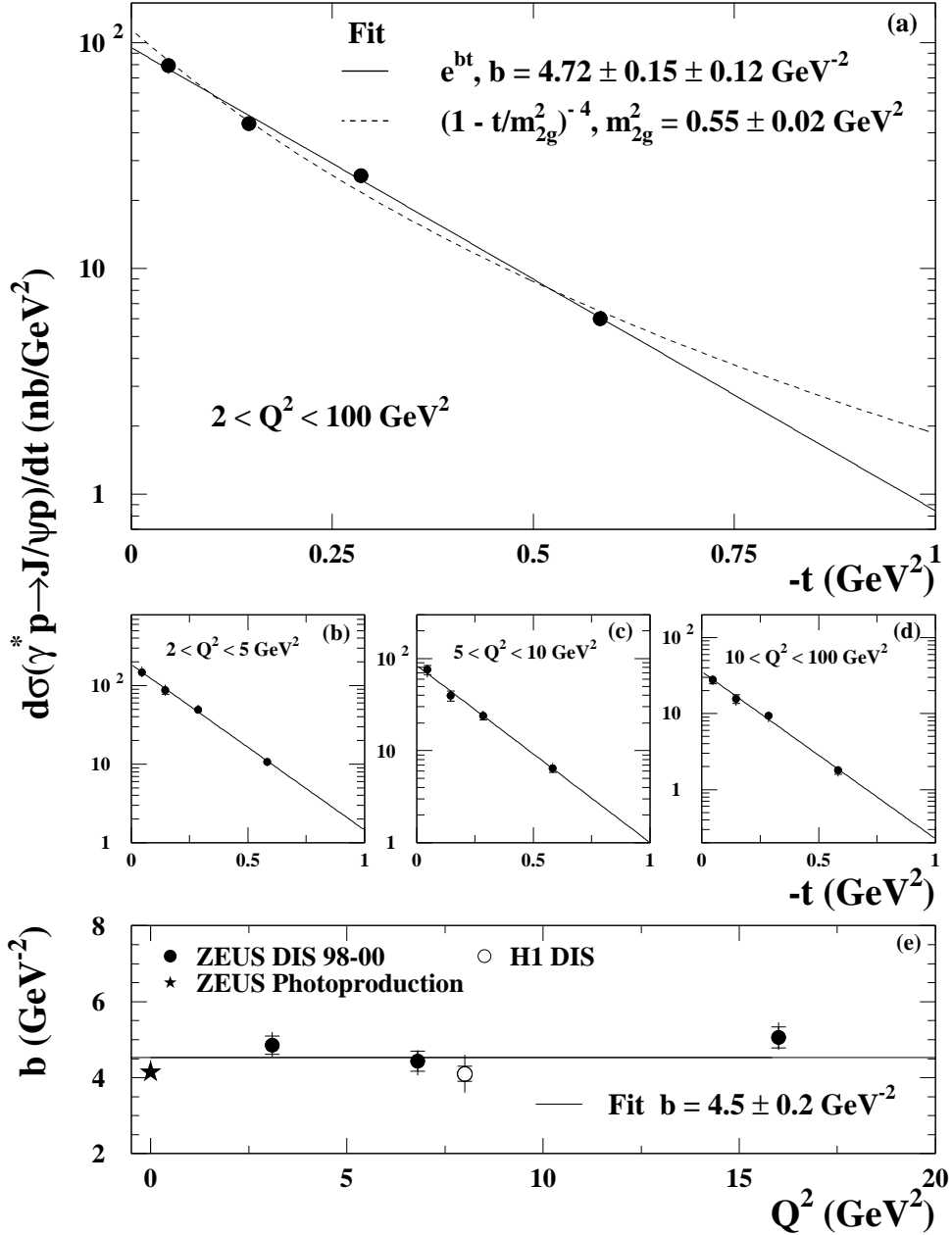,
width=13cm}
\end{center}
\caption{
Differential cross-sections $d\sigma /dt$ 
(a) over the entire $Q^2$ range and
(b)-(d) for three bins of $Q^2$, for $30<W<220\gev$ and $|t|<1\gev^2$. 
The full lines are the results of a fit to the form 
$d\sigma /dt = d\sigma /dt|_{t=0} \cdot e^{-b|t|}$ and
the dashed line is the result of a fit
using an elastic form factor assuming two-gluon exchange:
$d\sigma /dt \propto (1-t/m^2_{2g})^{-4}$. 
(e) The slope $b$, as a function of $Q^2$, 
compared to the ZEUS photoproduction and H1 results.
The mean value of $b$ is indicated by the horizontal line.
The inner error bars represent the statistical uncertainty,
the outer bars the statistical and systematic uncertainties 
added in quadrature. 
}
\label{fig-dsigmadt}
\vfill
\end{figure}

\begin{figure}[p]
\vfill
\begin{center}
\epsfig{file=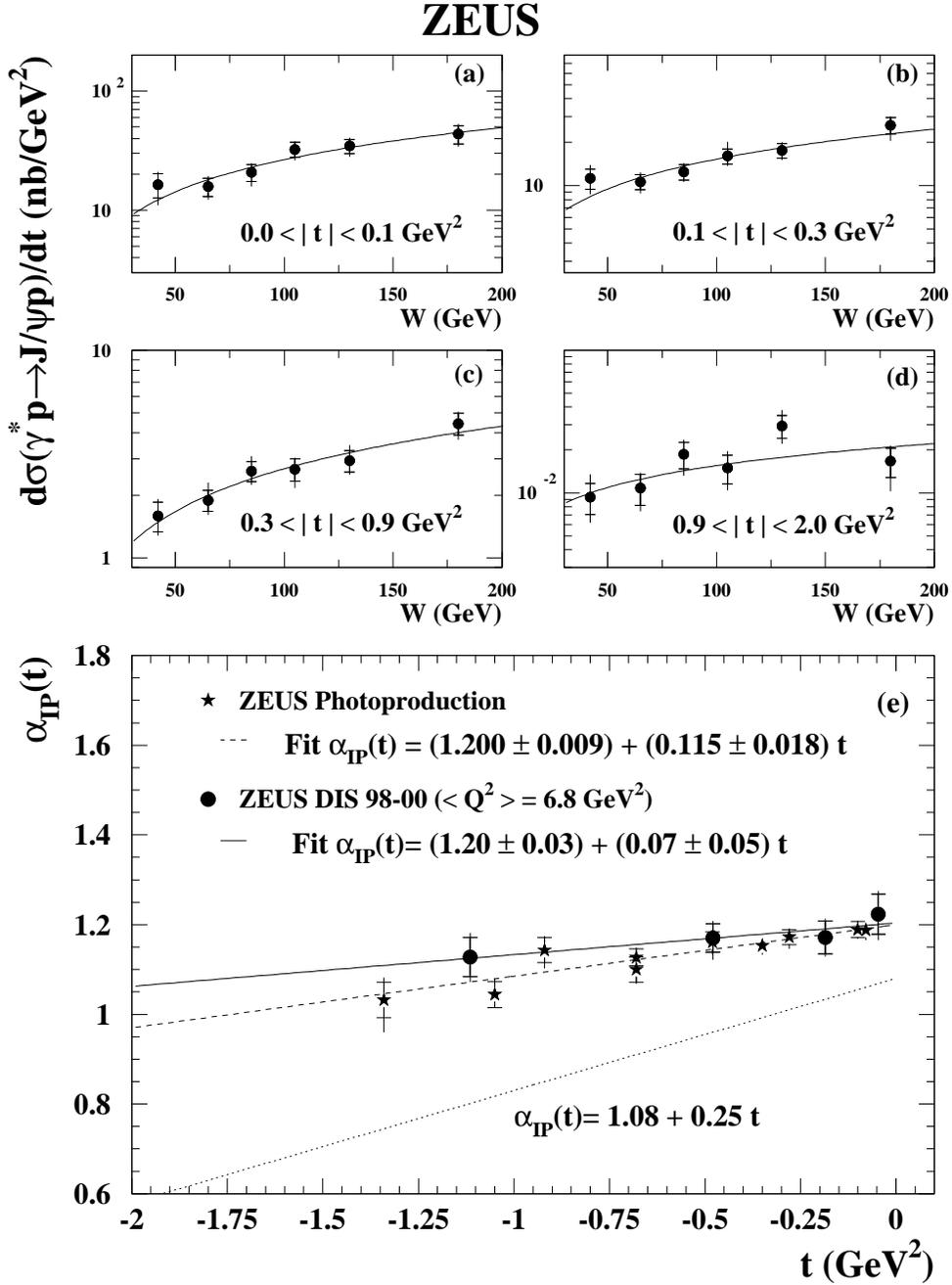,width=13cm}
\end{center}
\caption{
(a)-(d) Differential cross sections $d\sigma /dt$ 
as a function of $W$ for fixed ranges of $t$;
the full lines are fits to $W^{4(\alpha_{\pom}(t)-1)}$.
(e) Pomeron trajectory: 
the lines are linear fits to the data at $\langle Q^2\rangle=6.8\gev^2$ (full)
and to the $J/\psi$ photoproduction results (dashed);
the dotted line is the soft Pomeron trajectory~\protect\cite{np:b244:322}.
The inner error bars represent the statistical uncertainty, 
the outer bars the statistical and systematic uncertainties 
added in quadrature. 
}
\label{fig-regge}
\vfill
\end{figure}

\begin{figure}[p]
\vfill
\begin{center}
\epsfig{file=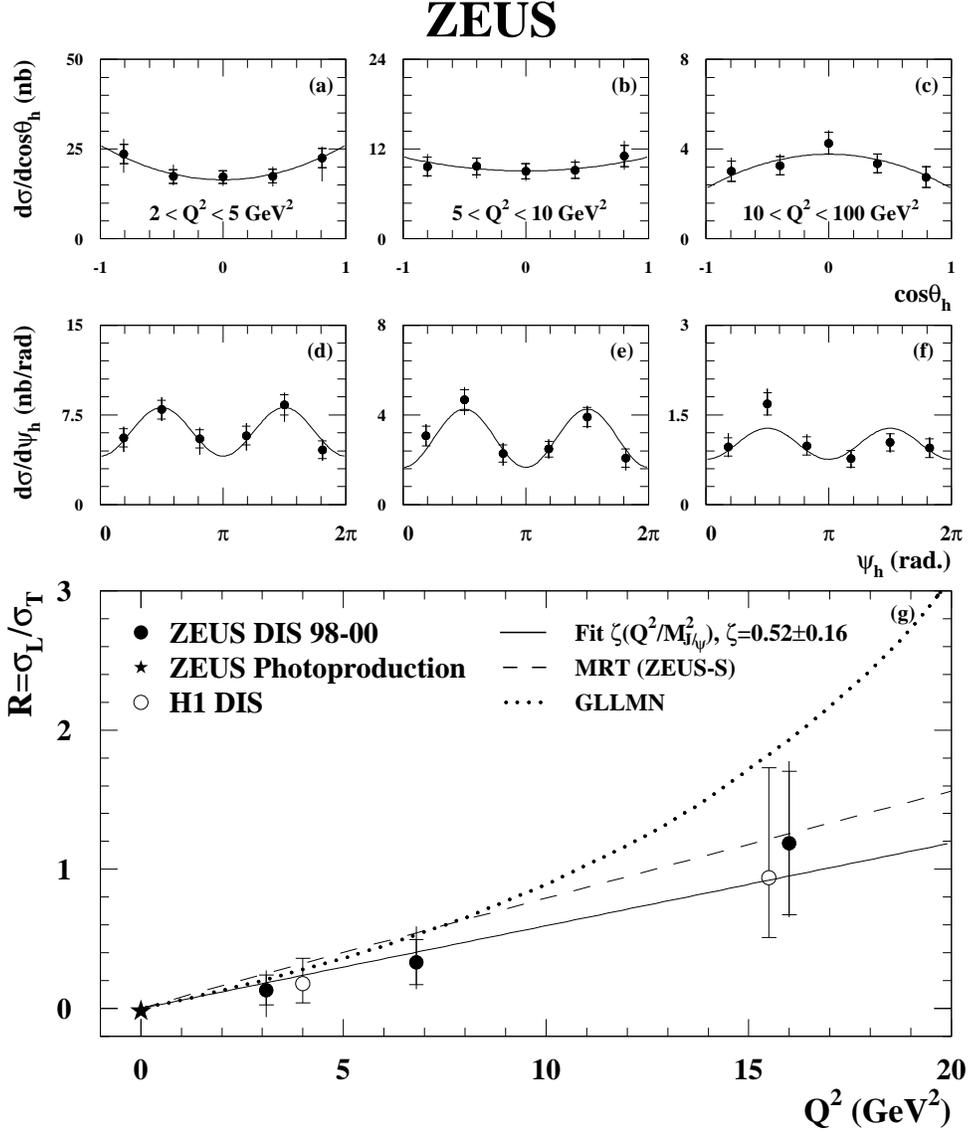,width=13cm}
\end{center}
\caption{
(a)-(f) Distributions of  $\cos \theta_h$ and $\Psi_h$ in three $Q^2$ bins; 
the curves are the fits  to Eqs.~(\ref{eq-angtheta}) and (\ref{eq-angpsi}). 
(g) Ratio $R=\sigma_L/\sigma_T$
as a function of $Q^2$;
the full curve is the result of the fit to the ZEUS data while
the dashed and dotted curves are the predictions of the MRT and GLLMN models, 
respectively.
The inner error bars represent the statistical uncertainty, 
the outer bars the statistical and systematic uncertainties 
added in quadrature.
}
\label{fig-angular}
\vfill
\end{figure}

\begin{figure}[p]
\vfill
\begin{center}
\epsfig{file=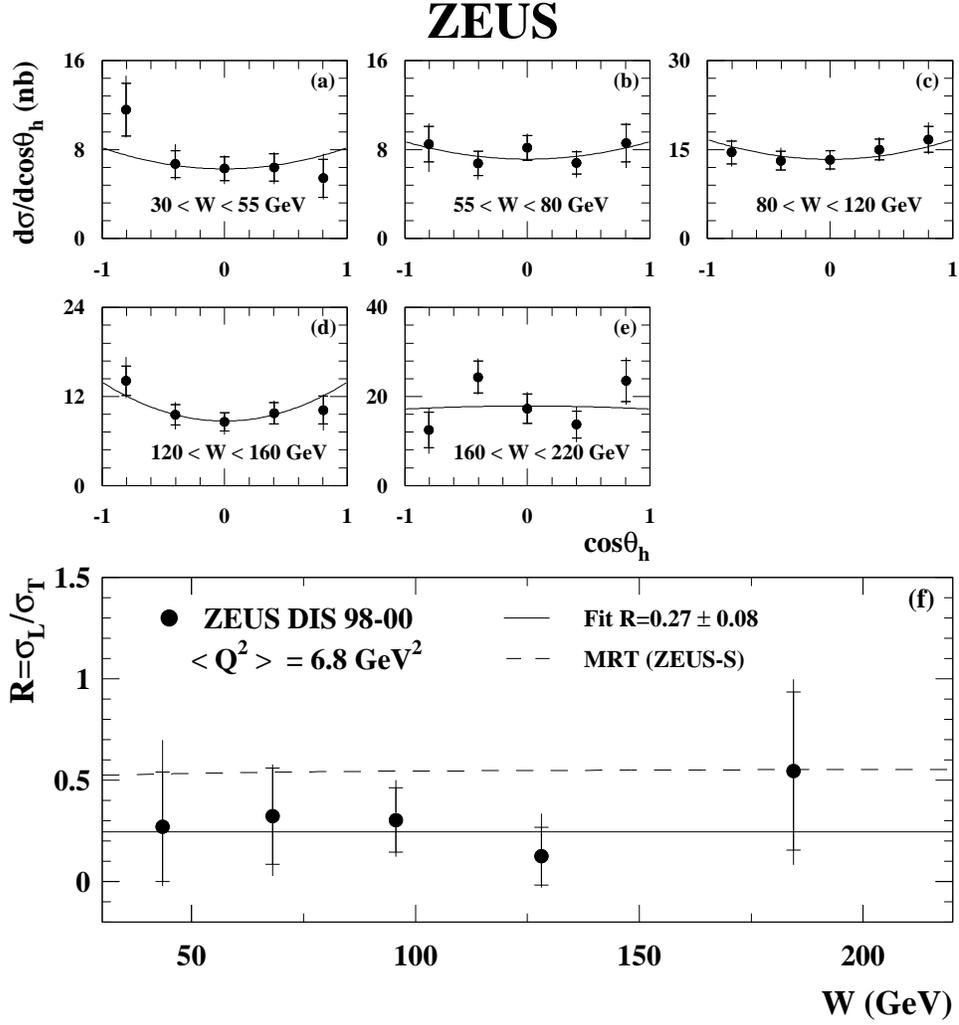,width=13cm}
\end{center}
\caption{
(a)-(e) Distributions of $\cos \theta_h$ for five bins of $W$;
the curves are the fits to Eq.~(\ref{eq-angtheta}). 
(f) Ratio $R=\sigma_L/\sigma_T$ as a function of $W$;
the dashed line is the MRT prediction and
the full line is the result of a one-parameter fit.
The error bars are statistical (inner) and total (outer). 
}
\label{fig-RvsW}
\vfill
\end{figure}

\begin{figure}[p]
\vfill
\begin{center}
\epsfig{file=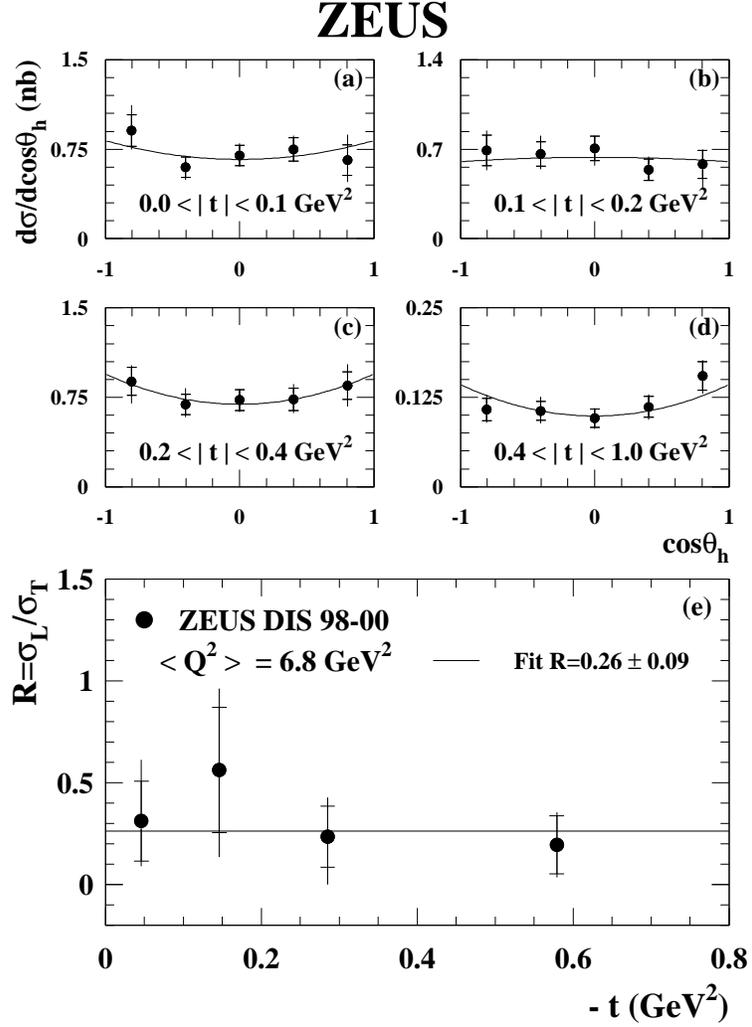,width=10cm}
\end{center}
\caption{
(a)-(e) Distributions of $\cos \theta_h$ for four bins of $t$;
the curves are the fits to Eq.~(\ref{eq-angtheta}). 
(f) Ratio $R=\sigma_L/\sigma_T$ as a function of $t$;
the full line is the result of a one-parameter fit.
The error bars are statistical (inner) and total (outer). 
}
\label{fig-Rvst}
\vfill
\end{figure}